%% file: neurips_2026.tex
\documentclass{article}

\PassOptionsToPackage{numbers, compress}{natbib}

\usepackage[preprint]{neurips_2026}

\usepackage[utf8]{inputenc} %
\usepackage[T1]{fontenc}    %
\usepackage{hyperref}       %
\usepackage{url}            %
\usepackage{booktabs}       %
\usepackage{amsfonts}       %
\usepackage{nicefrac}       %
\usepackage{microtype}      %
\usepackage{xcolor}         %
\usepackage{wrapfig}
\usepackage{graphicx}
\usepackage{amsmath}
\usepackage{subcaption}
\usepackage{tabularx}
\usepackage{multirow} 

\title{VocalGrad: Evaluating Acoustic Perception \\in Audio Language Models}

\author{
  Koki Ryu${}^{1, 2}$ \quad
  Hitomi Yanaka${}^{1, 2, 3}$ \\
  ${}^1$The University of Tokyo \\
  ${}^2$Riken \\
  ${}^3$Tohoku University \\
  \texttt{\{kokiryu, hyanaka\}@is.s.u-tokyo.ac.jp}}

\begin{document}

\maketitle

\begin{abstract}
Recent large audio language models (LALMs) have extended their capabilities beyond lexical speech understanding to tasks requiring paralinguistic and non-linguistic audio understanding, such as emotion recognition.
However, these models still struggle to reliably utilize acoustic information, particularly when it co-occurs with spoken linguistic content.
Existing benchmarks do not clearly distinguish acoustic perception from
higher-level reasoning, making it difficult to identify the source of these failures.
To address this issue, we introduce \textsc{VocalGrad}, a benchmark designed to evaluate LALMs' ability to perceive low-level acoustic variations.
VocalGrad formulates this problem as a simple binary classification of temporal change direction, enabling evaluation largely independent of higher-level reasoning or semantic understanding. 
Our benchmark results reveal a substantial gap in acoustic perception between human annotators and state-of-the-art LALMs: while human annotators achieve high accuracy, LALMs perform at near-chance levels.
We further conduct linear probing to identify structural bottlenecks in these models. Our analysis reveals two distinct failure modes: for some categories, task-relevant information is preserved in internal representations but not utilized by the language model head; for others, the information becomes progressively less linearly decodable across model layers.
Overall, our results reveal substantial limitations of LALMs in modeling acoustic variation under controlled settings and expose representation differences not observable from text outputs.
\end{abstract}

\input{chapters/1_introduction}
\input{chapters/2_related_work}
\input{chapters/3_dataset_construction}
\input{chapters/4_benchmarking}
\input{chapters/5_linear_probing}
\input{chapters/6_conclusion}

\begin{ack}
We thank the three anonymous reviewers and the meta-reviewer for their thoughtful engagement with our work and constructive feedback.
We also thank the human annotators for their time and effort in establishing the human performance baseline.
We are grateful to Yuki Saito, Kentaro Seki, Wataru Nakata, Shunsuke Kando, Amane Watahiki, Adam Nohejl, Zhihan Cao, Tomoki Doi, Gaëtan Margueritte, Xiaotian Wang, Ryoma Kumon, Taisei Yamamoto, and Kaito Watanabe for their valuable discussions and feedback on the manuscript.
This work was supported by JST CREST Grant Number JPMJCR2565, Japan.
\end{ack}

\newpage

\bibliographystyle{unsrtnat}
\bibliography{ref}

\newpage
\appendix
\input{chapters/appendix}

\newpage
\input{checklist.tex}

\end{document}

%% file: chapters/1_introduction.tex
\section{Introduction}
\label{sec:introduction}
\input{figures/vocalgrad-overview}

As Large Language Models (LLMs) advance, an increasing number of models can handle audio as an input modality, including both commercial~\citep{gemini1, gemini25, gpt-4o} and open-source~\citep{qwen35omni, kimi-audio, stepaudio2, audioflamingo} systems.
The capabilities of these large audio language models (LALMs) are no longer limited to recognizing spoken lexical content (i.e., automatic speech recognition, ASR), but also extend to tasks requiring the understanding of paralinguistic and non-linguistic information, such as emotion recognition and environmental sound understanding~\citep{mmsu, mmar}.
\par

However, recent work has shown that LALMs struggle to jointly utilize lexical content and acoustic cues.
For example, models tend to rely on lexical information when it conflicts with prosodic cues in emotion recognition~\citep{audio-listen}, fail to identify stressed words in speech~\citep{stresstest}, and exhibit limitations in jointly understanding speech and environmental sounds~\citep{joint-understanding}.
While these results highlight limitations in high-level tasks, they do not clarify whether the bottleneck arises from insufficient acoustic perception or from downstream integration with semantic representations.
\par

Existing benchmarks partially address the perception of primitive acoustic cues such as loudness or pitch~\citep{mmsu, mmar}.
However, their task formulations often involve additional reasoning components, such as reranking multiple audio clips or aligning audio events with absolute time references.
Under these settings, even when a model fails, it is difficult to determine whether the error originates from deficiencies in perceptual encoding or in subsequent reasoning.
\par

To address these issues, we introduce \textsc{VocalGrad}, a benchmark for evaluating the perception of acoustic transformations in speech.
In VocalGrad, we apply temporal transformations to a single acoustic feature within a speech clip and formulate the task as a binary classification problem that predicts whether the feature increases or decreases over time.
By preserving lexical content, isolating a single acoustic factor, and minimizing task complexity, our benchmark enables evaluation of perceptual ability largely independent of higher-level semantic understanding or reasoning.
This structured design also supports unified evaluation across both speaker-intrinsic factors (e.g., pitch, loudness) and environment-induced factors (e.g., noise, reverberation).
\par

Using this dataset, we evaluate multiple state-of-the-art LALMs and show that they struggle to solve even simple binary tasks involving basic acoustic changes, while human annotators achieve near-perfect accuracy.
Furthermore, we analyze the models' internal representations using linear probing to identify where acoustic information becomes undecodable.
This aspect cannot be revealed solely through text-based evaluation.
Our contributions are summarized as follows:
\begin{itemize}
    \item We propose \textsc{VocalGrad}\footnote{Code and dataset are available at public repositories: \\ \url{https://huggingface.co/datasets/ynklab/vocalgrad} \\
    \url{https://github.com/ynklab/vocalgrad}.}, a 21,600-clip benchmark that systematically isolates the ability of LALMs to perceive acoustic features.
    \item We evaluate five state-of-the-art LALMs against human annotators, showing that while the tasks are easy for humans, LALMs perform at a near-chance level.
    \item Through analysis of internal representations, we identify two distinct failure modes: for some transformations (e.g., volume, noise), information is preserved through transformer layers, but prediction is limited by the language model head; for other transformations, relevant information becomes progressively less linearly decodable across language decoder layers.
    \item We further show that fine-tuning on specific acoustic transformations induces cross-category changes in internal representations and performance.
\end{itemize}

%% file: figures/vocalgrad-overview.tex
\begin{figure}[t]
    \centering
    \includegraphics[width=\linewidth]{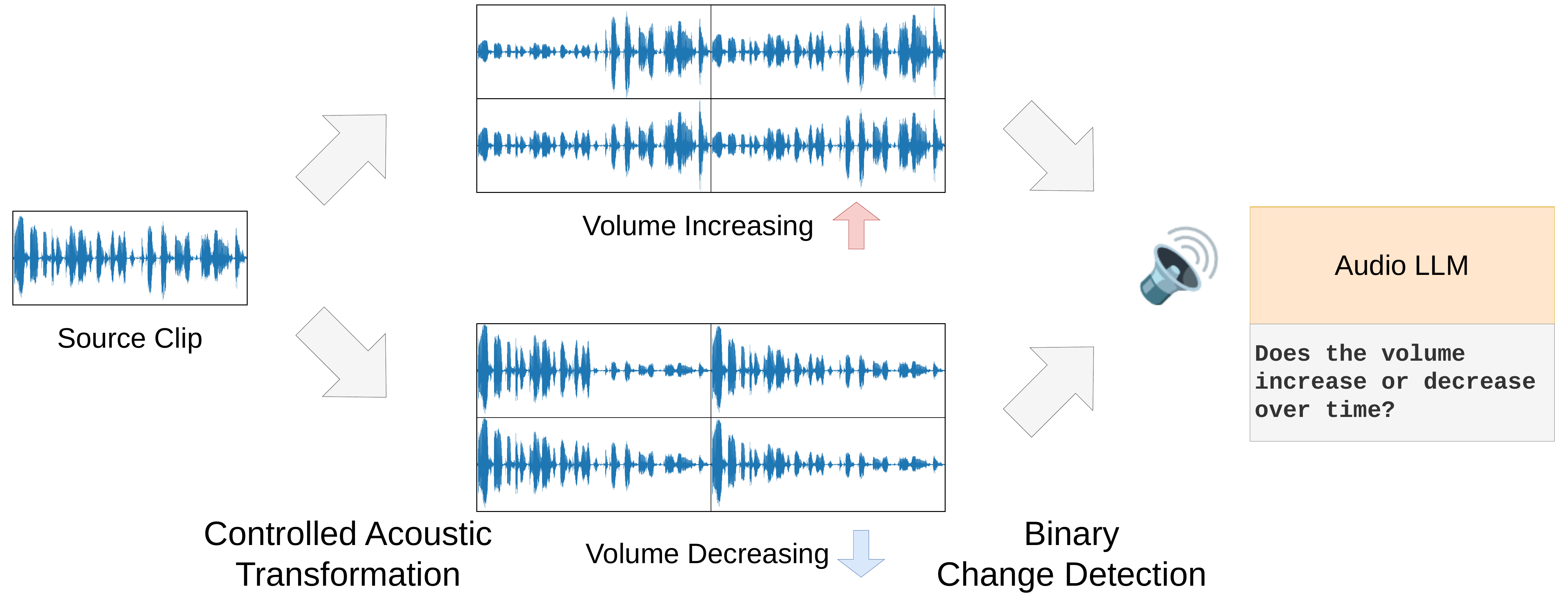}
    \caption{
Overview of VocalGrad.
Each source speech clip is transformed with a temporal change applied to a single acoustic factor.
The task is formulated as a binary classification of change direction, enabling evaluation of acoustic perception in isolation from semantic understanding required for downstream tasks.
    }
    \label{fig:vocalgrad-overview}
\end{figure}

%% file: chapters/2_related_work.tex
\section{Related Work}
\label{sec:related_work}

\subsection{Benchmarks for Acoustic Perception}

Recent audio understanding benchmarks include tasks that require models to process paralinguistic and non-linguistic information in audio signals~\citep{mmau, mmsu} and are widely used in technical reports of modern LALMs~\citep{qwen35omni, kimi-audio}.
\par

However, several studies have shown that even state-of-the-art models struggle to fully utilize such information.
For example, LISTEN~\citep{audio-listen} demonstrates that LALMs tend to over-rely on lexical content when prosodic cues contradict the literal meaning of speech.
Similarly, StressTest~\citep{stresstest} shows that tasks such as stressed-word identification and intent inference remain challenging without task-specific fine-tuning.
\par

Related limitations have also been observed for non-linguistic acoustic factors.
\citet{silence-matters} show that background noise can negatively affect reasoning performance, while \citet{joint-understanding} report that LALMs struggle to jointly process speech and environmental sounds.
These findings suggest that current models fail to reliably leverage acoustic information, but do not clarify where in the processing pipeline these failures arise.
\par

Existing benchmarks partially include tasks related to acoustic perception.
For instance, MMSU~\citep{mmsu} includes sorting tasks based on acoustic properties across multiple speech segments but requires additional reasoning steps, such as cross-segment comparisons and reranking.
WoW-Bench~\citep{wow-bench} evaluates LALMs' acoustic perception
using marine mammal vocalizations, but its tasks involve
additional reasoning, such as cross-segment comparisons and
interpreting detailed temporal acoustic patterns.
MMAU~\citep{mmau} includes detection tasks for acoustic changes, yet requires answers in the form of timestamps, meaning that correct perception alone does not directly lead to correct predictions.
SALMon~\citep{salmon} considers acoustic condition changes within audio clips, but focuses on high-level attributes such as sentiment or speaker identity and relies on multi-clip likelihood-based evaluation.
\par

In contrast, VocalGrad applies transformations to a single acoustic factor and formulates the task as a simple binary classification of temporal change direction.
This design enables direct evaluation of acoustic perception, largely independent of downstream reasoning processes.

\subsection{Analysis of Internal Representations in Audio Models}

Internal representation analysis of audio models has primarily focused on models trained with self-supervised learning objectives.
For speech representation models such as HuBERT~\citep{hubert} and wav2vec 2.0~\citep{wav2vec2}, prior work has examined how acoustic and semantic features are encoded across layers.

For example, \citet{lexical-tone-slm} show that such models strongly encode acoustic properties such as lexical tone, while \citet{phonetic-speech} demonstrate that embeddings reflect phonetic similarity.
Furthermore, \citet{phonetic-phonemic} report that both phonetic and phonemic information can be extracted from intermediate representations with high accuracy.
\citet{speech-encoder-probe} conduct large-scale layer-wise probing of speaker-specific acoustic, prosodic, and paralinguistic attributes across multiple models and reveal attribute-dependent patterns in how such information is encoded across layers.
\par

More recent work extends this analysis to LALMs, which combine speech encoders with LLM modules.
These studies investigate how information is transferred from the audio encoder to downstream language components.
For example, \citet{inner-repr-slm} analyze the alignment between speech inputs and internal representations, while \citet{beyond-transcription} and \citet{audiolens} focus on how higher-level acoustic attributes, such as speaker identity and emotion, are represented.
\par
While these studies provide important insights into the internal mechanisms of LALMs, they primarily focus on relatively high-level attributes.
In contrast, our work focuses on controlled temporal changes in lower-level acoustic features.
In addition, we trace task-relevant information across the LALM processing pipeline, from pre-LLM audio representations through language decoder layers to text-prediction representations, rather than focusing solely on representations within the language decoder.
This allows us to distinguish between the degradation of task-relevant acoustic information and failures to utilize information that remains encoded in late LLM representations.

%% file: chapters/3_dataset_construction.tex
\section{Dataset Construction}
\label{sec:dataset_construction}

\subsection{Overview}
\input{tables/vocalgrad_statistics}

Table~\ref{tab:vocalgrad_statistics} summarizes key statistics of the VocalGrad dataset.
It contains 21,600 audio clips across nine categories of acoustic transformations, including three speaker-intrinsic transformations and six environment-induced transformations.
Details of each transformation are provided in Section~\ref{sec:transformation-categories}.
\par

Each sample is labeled with one of three intensity tiers and one of four temporal trajectories, enabling fine-grained analysis of model behavior.
The benchmark is formulated as a binary classification task that predicts the direction of temporal change (increasing vs.\ decreasing).
Each category contains an equal number of increasing and decreasing samples.
This controlled and symmetric design facilitates systematic analysis using various methods, including linear probing.

\subsection{Common Pipelines}
\label{sec:dataset-common-pipeline}

\paragraph{Source Clip Selection}
We select source speech clips from the VCTK dataset~\citep{vctk}, which contains recordings of speakers with diverse accents reading scripted sentences such as newspaper articles.
We first select 50 speakers, balancing gender and accent.
We then filter out clips shorter than three seconds and randomly choose two clips per speaker, resulting in 100 source clips.
The same set of source clips is used across all categories, and each source clip yields 24 augmented samples, resulting in 2,400 samples per category.
In Appendix~\ref{sec:different-sources}, we additionally apply the same data augmentation and benchmarking pipeline to different source datasets and examine whether the observed performance trends persist across source domains.

\paragraph{Speech Onset and Offset Detection}
Some transformations are difficult to perceive in non-speech regions.
To ensure that transformations are applied to speech segments, we estimate speech onset and offset using a simple energy-based heuristic~\citep{onset-detection}.
Specifically, we compute short-time RMS over fixed windows, estimate a noise floor from the initial portion of each clip, and mark frames as speech when their RMS exceeds an adaptive threshold.
To suppress spurious detections, only sequences of consecutive active frames longer than a predefined minimum length are retained.
The detected onset and offset are then manually verified using waveform visualizations.

\paragraph{Data Augmentation}
For each category, unless otherwise specified, the detected speech span is divided into 0.05-second segments.
Transformations are applied to each segment with varying parameters to simulate temporal changes.
To ensure smooth transitions, we introduce a 0.005-second crossfade between adjacent segments where applicable.
\par

Each source clip is augmented along three dimensions:
(1) direction of change (increasing or decreasing),
(2) intensity tiers (high, medium, low), and
(3) temporal trajectories, including jump (abrupt change), linear, and quadratic variants (quad\_first\_flat and quad\_last\_flat).
All combinations are applied to each clip, resulting in 24 augmented samples per source clip.
This augmentation procedure is consistent across categories, enabling unified evaluation of LALMs on diverse temporal acoustic variations.
Detailed definitions of the temporal trajectories are provided in Appendix~\ref{sec:temporal-trajectory-details}.

\subsection{Category-Specific Augmentations}
\label{sec:transformation-categories}

This section outlines the augmentation procedures used to generate each category in VocalGrad.
We consider three speaker-intrinsic transformations (Speaking Speed, Voice Pitch, and Volume) and six environment-induced transformations (Audio Distortion, Audio Roughness, Background Noise, Echo, Voice Clarity, and Voice Vibration), covering a diverse range of acoustic effects.
Detailed hyperparameter settings are provided in Appendix~\ref{sec:dataset-hyperparameters}.

\begin{itemize}
    \item \textbf{Speaking Speed (\textit{speed})} Modify the speaking rate using librosa~\citep{librosa}.
    \item \textbf{Voice Pitch (\textit{pitch})} Shift the pitch using librosa~\citep{librosa}.
    \item \textbf{Volume (\textit{volume})} Adjust loudness via gain control.
    \item \textbf{Audio Distortion (\textit{distortion})} Downsample audio to simulate low-resolution distortion~\citep{distortion-literature}.
    \item \textbf{Audio Roughness (\textit{roughness})} Apply bit-depth reduction to introduce quantization artifacts.
    \item \textbf{Background Noise (\textit{noise})} Add Gaussian noise to the signal.
    \item \textbf{Echo (\textit{echo})} Apply a reverberation-based echo effect using a Schroeder reverberator~\citep{reverbation}.
    \item \textbf{Voice Clarity (\textit{clarity})} Reduce high-frequency components using low-pass filtering.
    \item \textbf{Voice Vibration (\textit{vibration})} Apply low-frequency amplitude modulation.
\end{itemize}

\subsection{Prompting Format}
\label{sec:prompting-format}

Each clip is paired with a binary question using a shared prompt template:
\textit{Does the \{attribute\} increase or decrease over time? Answer with only one word: ``increase'' or ``decrease''.}
Using a common prompt across all categories allows us to compare model behavior under a consistent evaluation setting while varying only the acoustic transformation type.
We further analyze the sensitivity of LALMs to prompt variations in Appendix~\ref{sec:prompt-sensitivity}.
We additionally evaluate alternative problem formulations that combine the augmented clips with the original source clips, including binary change detection and ternary classification, in Appendix~\ref{sec:different-problem-formats}.
\par

Because the prompt explicitly refers to the transformation attribute, correct answers require the model to associate each attribute name with its corresponding acoustic phenomenon.
We verify in Section~\ref{sec:benchmarking-results} that human annotators can reliably make this association and answer the task correctly.

%% file: tables/vocalgrad_statistics.tex
\begin{table}[t]
\centering
\caption{Key statistics of VocalGrad.}
\label{tab:vocalgrad_statistics}
\begin{tabular}{@{}lclc@{}}
\toprule
\textbf{Statistics} & \textbf{Value} & \textbf{Statistics} & \textbf{Value} \\
\midrule
Total Questions & 21,600 & Intensity Tiers & 3 \\
Speaker-Intrinsic Categories & 3 & Transformation Trajectories & 4 \\
Environment-Induced Categories & 6 & Avg. Audio Length & 3.98 s \\
\bottomrule
\end{tabular}
\end{table}

%% file: chapters/4_benchmarking.tex
\section{Benchmarking}
\label{sec:benchmarking}

\subsection{Evaluation Settings}
\label{sec:evaluation-settings}

We benchmark state-of-the-art LALMs using VocalGrad.
We evaluate both commercial and open-source models, including Gemini 3 Flash\footnote{https://deepmind.google/models/gemini/flash/}, Kimi-Audio~\citep{kimi-audio}, MiMo-Audio~\citep{mimo-audio}, Step-Audio~2~mini~\citep{stepaudio2}, and Audio~Flamingo~3~\citep{audioflamingo}, selected based on their strong performance on existing benchmarks~\citep{mmau, mmsu}.
These models differ in their acoustic modeling strategies, ranging from autoregressive token modeling to text-aligned supervision (see Appendix~\ref{sec:acoustic-modeling} for details).
Runtime configurations are provided in Appendix~\ref{sec:model-runtime-parameter}.
\par
To ensure consistent evaluation across models, we adopt accuracy on text outputs as the primary metric, since raw log probabilities are unavailable for certain models, such as Gemini 3 Flash.
Since some model outputs do not strictly follow the required format, we use case-insensitive, whole-word matching of “increase”/“increases” and “decrease”/“decreases”, accepting only outputs that contain exactly one of the two directions.
We analyze the effect of this parsing strategy in Appendix~\ref{sec:parsing-failure-analysis}.
\par

To compare model performance with human perception, we conduct a human evaluation study with six annotators regularly exposed to English speech.
For each category, we sample 50 clips and collect binary annotations (increase vs.\ decrease) from at least three annotators per clip.
The sampled clips share the same base VCTK sources across categories, with approximately balanced coverage of intensity tiers and temporal trajectories.
Annotations are collected through a dedicated web interface, and we report the average accuracy across all responses.
Details of the sampling procedure, its limitations, and the annotation interface are provided in Appendix~\ref{sec:human-annotation-tool}.

\subsection{Results}
\label{sec:benchmarking-results}
\input{tables/human_model_accuracy_comparison}

Table~\ref{tab:vocalgrad-human-model-accuracy} reports the classification accuracy of human annotators and LALMs on VocalGrad.
We additionally provide performance breakdowns by temporal trajectory and intensity tier in Appendix~\ref{sec:temporal-trajectory-analysis} and Appendix~\ref{sec:intensity-tier-analysis}, respectively.
As shown in the table, human annotators achieve an average accuracy of 94.4\% across categories, reaching 100\% in several cases.
In contrast, all evaluated LALMs perform at a near-chance level across categories.
Even the best-performing model, Kimi-Audio, achieves only 51.8\% on the overall average.
The models also remain near chance under alternative prompt formulations
(Appendix~\ref{sec:prompt-sensitivity}), suggesting that their poor performance
cannot be explained solely by prompt wording.
\par
These results indicate that even acoustic changes easily perceived by humans remain challenging for current LALMs when evaluated in a text-based binary classification setting.
A detailed analysis of failure patterns is provided in Appendix~\ref{sec:benchmark-failure-analysis}.

\subsection{Ablation: Removing Linguistic Content}
\label{sec:ablation-beep}
\input{tables/ablation_beep_table}

To better understand whether the observed failures are due to linguistic complexity or limitations in acoustic perception, we conduct an ablation study using synthetic sine-wave signals rather than speech clips.
Since several transformations depend on speech-specific temporal or spectral structure, they are difficult to perceive in pure sine-wave signals, which lack such structure.
Accordingly, we restrict this experiment to Voice Pitch, Volume, Background Noise, and Voice Vibration.
\par
As this experiment is intended as a controlled ablation of linguistic content, we use a small set of 10 base clips consisting of pure sine waves with different constant frequencies.
Each clip has a duration of 3.6 seconds, with the first and last 0.75 seconds set to silence to match the temporal structure of the VCTK-based clips.
We apply the same VocalGrad transformation pipeline to these signals, generating 240 samples per category.
These clips are evaluated using the same benchmarking protocol as the main dataset.
\par

Table~\ref{tab:vocalgrad-ablation-beep-accuracy} reports model performance on this ablation dataset.
Compared to the original VocalGrad results, several models, including Kimi-Audio and MiMo-Audio, show improved accuracy in this simplified setting.
However, none of the models exceed 70\% accuracy in any category, despite the binary nature of the task.
These results suggest that removing linguistic content in speech clips does not fully resolve the challenge.
Even in the absence of linguistic variability, LALMs still struggle to reliably detect simple acoustic transformations.
This indicates that the limitation of current LALMs is not solely due to interference from linguistic structure, but reflects a more fundamental deficiency in modeling temporal acoustic variation.

%% file: tables/human_model_accuracy_comparison.tex
\begin{table}[t]
\centering
\small
\setlength{\tabcolsep}{2pt}
\caption{Human annotator average and model accuracies (\%) on VocalGrad by category.}
\label{tab:vocalgrad-human-model-accuracy}
\begin{tabular}{lrrrrrrrrrr}
\toprule
Model & Avg. & speed & pitch & volume & distortion & roughness & noise & echo & clarity & vibration \\
\midrule
Human Avg. & 94.4 & 95.3 & 84.7 & 100.0 & 86.0 & 99.3 & 100.0 & 94.7 & 90.7 & 99.3 \\
\midrule
Gemini 3 Flash & 50.2 & 50.0 & \textbf{51.5} & 49.9 & 48.1 & 44.8 & 55.5 & 50.3 & \textbf{54.0} & 47.6 \\
Kimi-Audio & \textbf{51.8} & \textbf{53.1} & 50.5 & 50.5 & \textbf{51.6} & 48.0 & \textbf{56.5} & \textbf{54.5} & 53.6 & 47.6 \\
MiMo-Audio & 50.3 & 50.7 & 50.5 & \textbf{52.9} & 49.8 & 43.2 & 53.2 & 50.0 & 52.2 & \textbf{50.3} \\
Step-Audio~2~mini & 50.0 & 49.9 & 50.2 & 49.9 & 50.2 & \textbf{49.9} & 49.9 & 50.0 & 49.8 & 49.8 \\
Audio~Flamingo~3 & 49.6 & 50.0 & 50.0 & 50.0 & 50.0 & 46.7 & 50.0 & 50.0 & 49.9 & 50.0 \\
\bottomrule
\end{tabular}
\end{table}

%% file: tables/ablation_beep_table.tex
\begin{table}[t]
\centering
\small
\caption{Model accuracies (\%) on VocalGrad ablation-beep subsets by category.}
\label{tab:vocalgrad-ablation-beep-accuracy}
\begin{tabular}{lrrrr}
\toprule
Model & pitch & volume & noise & vibration \\
\midrule
Gemini 3 Flash & 48.3 & 50.0 & 47.5 & 49.2 \\
Kimi-Audio & \textbf{50.4} & \textbf{57.5} & 50.0 & 50.0 \\
MiMo-Audio & 50.0 & 55.8 & \textbf{64.2} & 50.0 \\
Step-Audio~2~mini & 50.0 & 50.0 & 50.0 & 50.0 \\
Audio~Flamingo~3 & 50.0 & 50.0 & 50.0 & 50.0 \\
\bottomrule
\end{tabular}
\end{table}

%% file: chapters/5_linear_probing.tex
\section{Linear Probing}
\label{sec:linear_probing}

\subsection{Method Overview}
\label{sec:linear-probe-method}
In this section, we take a further step toward identifying the bottlenecks in LALM performance on VocalGrad.
Specifically, we use linear probing to quantify the amount of task-relevant information encoded in hidden representations.
Given an audio clip and its binary label (increase vs.\ decrease), we train a simple linear classifier on top of each extracted representation.
We then compare classification accuracy across different representations and layers to identify how the linear accessibility of task-relevant information changes within the model.
\par
For the main analysis, we consider two types of representations:
Audio Representations and Text Prediction Representations, as described below.
While the main analysis uses temporally aggregated audio representations and
the output-prediction position, we additionally examine how task-relevant
information is distributed across individual token positions in
Appendix~\ref{sec:position-wise-probing}.

\paragraph{Audio Representations}
We first analyze representations aligned with the input audio tokens.
These representations allow us to examine how acoustic information is transformed through the model pipeline.
Specifically, we consider the following three types of representations:

\begin{itemize}
\item \textit{Mel}: The mel-spectrogram extracted directly from the raw audio input. This serves as a reference representation containing the original acoustic information and is shared across all models.
\item \textit{Pre-LLM}: Audio representations obtained after processing by components such as an audio encoder or tokenizer, but before the language decoder layers.
\item \textit{Late-LLM}: The representation from the final layer of the language decoder, capturing the effect of subsequent transformations within the LLM.
\end{itemize}

The goal of probing these representations is to assess how much information relevant to the VocalGrad task remains at each stage of processing.
To handle variable-length inputs, we divide each audio clip into 16 temporal segments and apply average pooling within each segment.
The pooled segment representations are then concatenated to form a fixed-length input for the linear classifier.
\par

\paragraph{Text Prediction Representations}
We further investigate how temporal acoustic information is integrated into representations used for output token prediction.
Specifically, we extract hidden representations corresponding to the output-prediction position across all layers of the language decoder.
By applying linear probing to these representations, we analyze how task-relevant information evolves across layers and identify where it becomes undecodable.

\subsection{Experimental Setup}
To train the linear classifiers, we construct a separate training set for VocalGrad.
Specifically, we select 57 speakers from the VCTK dataset, ensuring that they do not overlap with the speakers used in the test set.
For each speaker, we choose 8 or 9 clips whose lexical content does not overlap with the test data, resulting in 500 base clips in total.
We then apply the same augmentation pipeline described in Section~\ref{sec:dataset-common-pipeline} to these source clips.
As a result, we generate 12,000 training samples for each transformation category.
\par

The linear classifier is implemented as a single linear layer in PyTorch~\citep{pytorch}, trained using the AdamW optimizer~\citep{adamw}.
Target representations are extracted from the four open-source models evaluated in Section~\ref{sec:benchmarking}.
Further implementation details are provided in Appendix~\ref{sec:model-runtime-parameter}.

\subsection{Results}
\label{sec:linear_probing_results}

\input{figures/category-accuracy-by-audio-stage}
\input{figures/linear_probe_text}

Figure~\ref{fig:category-accuracy-by-audio-stage} and Figure~\ref{fig:linear-probe-text} show the linear probing accuracy on audio representations and text prediction representations, respectively.

First, the results on the audio representations indicate that the amount of task-relevant information preserved in the representations depends on both the model and the type of acoustic transformation.
Overall, representations derived from Kimi-Audio and MiMo-Audio tend to achieve higher probing accuracy compared to those from other models.
One possible explanation is that these models employ autoregressive audio modeling objectives, which may encourage richer temporal representations in their audio tokens.
An exception is the Speaking Speed category, where MiMo-Audio consistently performs worse.
Figure~\ref{fig:category-accuracy-by-audio-stage} suggests that this difference already emerges in the pre-LLM representations.
One possible interpretation is that MiMo-Audio's tokenization process encodes spoken-word density information in a form less accessible through average pooling and linear probing than the representations produced by the other models' audio frontends.
Further investigation is needed to clarify the underlying cause of this difference.
\par

The results on the text prediction representations indicate that task-relevant information about temporal acoustic changes is aggregated at very early layers of the language decoder (around layer 5).
Beyond this point, the information either remains preserved or gradually attenuates as it propagates through deeper layers.
This observation suggests that although the models can extract relevant acoustic information at early stages, they often fail to reliably retain or utilize it in the final output layer.
One possible explanation is that current training objectives do not sufficiently encourage the retention and use of low-level acoustic information throughout the language decoder.
\par

Consistent with this observation, probing the Late-LLM audio representations reveals that accuracy falls below the Pre-LLM level in more than half of the categories, indicating that task-relevant information becomes less linearly accessible as it propagates through the transformer layers.
However, for Volume, Background Noise, and Audio Roughness, classifiers trained on Late-LLM representations still achieve near-perfect accuracy.
A similar pattern is observed in the near-final text prediction representations.
\par

This suggests that for certain acoustic transformations, LALMs preserve information in a form that remains linearly decodable even after passing through the language model layers.
In contrast, for the remaining categories, task-relevant information becomes
progressively less linearly decodable toward the output layer.
Therefore, for these categories, the low end-to-end performance observed in
Section~\ref{sec:benchmarking-results} cannot be explained solely
by a failure to read out information preserved in late-layer representations.
\par
We hypothesize that these category-dependent differences in information retention may be related to the temporal characteristics of the acoustic cues required for each transformation.
Transformations that remain decodable at later layers, such as Volume and Background Noise, provide relatively stable local cues, such as short-time energy, that can be estimated from short temporal windows.
In contrast, some transformations with low late-LLM probing accuracy require integrating information over longer temporal contexts, despite being readily perceived by humans.
For example, determining the direction of Speaking Speed requires comparing the density and duration of speech events across larger portions of the clip, while determining the direction of Voice Vibration requires tracking how an amplitude-modulation pattern evolves across multiple modulation cycles.
Such temporally distributed cues may be more vulnerable to temporal compression or attenuation as they propagate through the language decoder than frame-local cues.
We emphasize that this interpretation remains a hypothesis, and leave a more detailed investigation for future work.

\subsection{LoRA Fine-tuning}
\label{sec:fine-tuning}
To investigate whether the insights from the probing analysis translate into improvements in end-to-end performance, we perform LoRA~\citep{lora} fine-tuning on Kimi-Audio, the best-performing model in Section~\ref{sec:benchmarking}.
Results for the other models are reported in Appendix~\ref{sec:fine-tuning-all}.

\paragraph{Method}
Based on the observation that task-relevant information is preserved in the final layers for certain categories, we construct two fine-tuning variants:
(1) applying LoRA only to the parameters of the language model head (LM-Head), and
(2) applying LoRA to all linear layers of the model, including attention projections, feedforward layers, and the language model head (All-Linear).

Each model is trained using the category-specific training data described in the linear probing setup.
Detailed configurations are provided in Appendix~\ref{sec:model-runtime-parameter}.
\par

\paragraph{Results}
\input{tables/fine_tune_accuracy_table}
\input{figures/fine_tuning_visualization}

Table~\ref{tab:fine_tune_accuracy} shows the performance of each fine-tuned model on its corresponding test category.

For LM-Head fine-tuning, the model achieves high accuracy (88--99\%) on Volume, Background Noise, and Audio Roughness.
This supports the findings in Section~\ref{sec:linear_probing_results}, indicating that for these categories, a major bottleneck lies in the language model head.

With All-Linear fine-tuning, the model achieves near-perfect accuracy in most categories, suggesting that the remaining performance gap can be addressed by updating earlier layers.

We further evaluate cross-category generalization by applying models trained on one category to other categories.
The results are shown in Figure~\ref{fig:kimi-ft-heatmap}.
Notably, models fine-tuned on a single category often improve performance on related categories; for example, the fine-tuned model trained on Audio Distortion shows improved performance on Audio Roughness and Background Noise.
In some cases, accuracies significantly below chance are also observed, indicating that the model has learned systematic but inverted decision boundaries.
\par

These results may be partially explained by similarities between augmentation types, but they also raise the possibility that fine-tuning induces representations that are shared across multiple types of temporal acoustic change.
Further analysis is required to confirm this hypothesis.
\par
In Appendix~\ref{sec:emotion-recognition}, we further examine how VocalGrad fine-tuning affects downstream emotion recognition on natural conversational speech.

\paragraph{Linear Probing on Tuned Models}

We also perform linear probing on the fine-tuned models using the same protocol as in Section~\ref{sec:linear-probe-method}.
Results for models fine-tuned on Volume and Voice Pitch are presented in Figure~\ref{fig:kimi-ft-probing}.
\par

Overall, probing accuracy improves compared to the base model.
In some cases, such as the Voice-Pitch-tuned model evaluated on Speaking Speed, probing accuracy on unseen categories reaches near-perfect levels at the final layer.
\par

However, these improvements in linear separability are not always reflected in end-to-end task accuracy (Figure~\ref{fig:kimi-ft-heatmap}).
This suggests that category-specific fine-tuning can induce representational changes that extend to unseen categories, even when these changes are not directly reflected in textual predictions.
We further investigate whether these cross-category effects reflect the emergence of shared representational structures across categories in Appendix~\ref{sec:fine-tuning-category-alignment}.

%% file: figures/category-accuracy-by-audio-stage.tex
\begin{figure}[t]
\centering
\includegraphics[width=\linewidth]{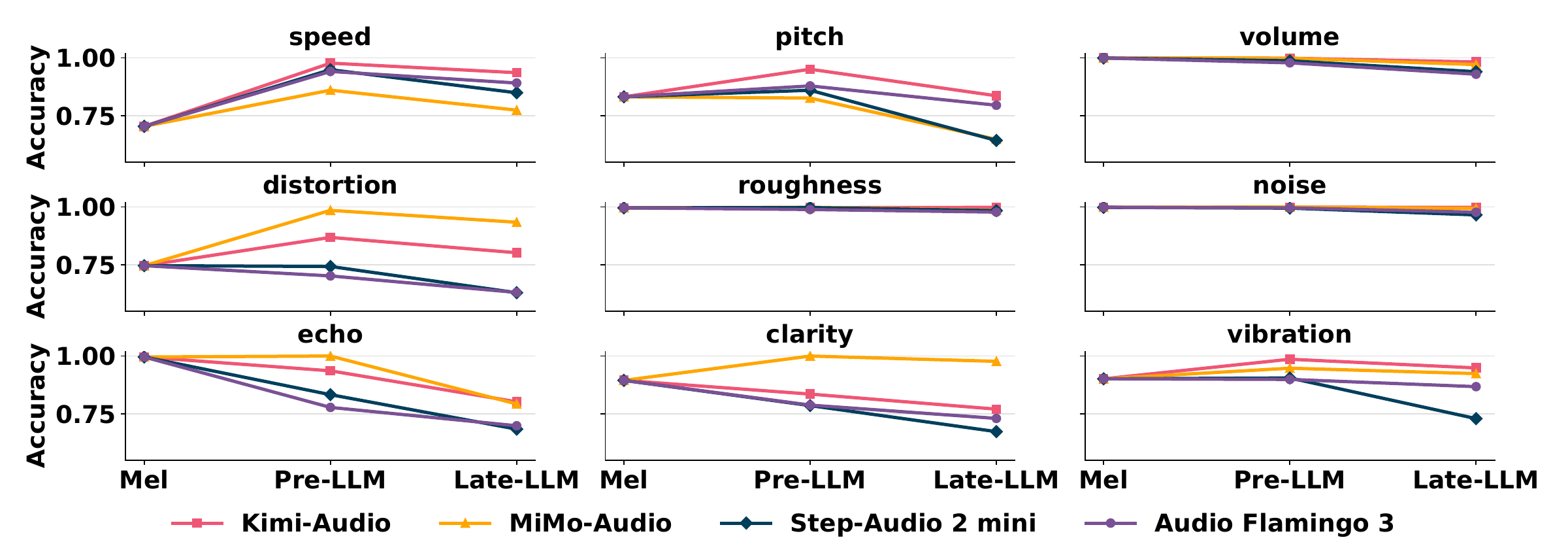}
\caption{Linear probing accuracies with different audio representations.}
\label{fig:category-accuracy-by-audio-stage}
\end{figure}

%% file: figures/linear_probe_text.tex
\begin{figure}[t]
\centering
\includegraphics[width=\linewidth]{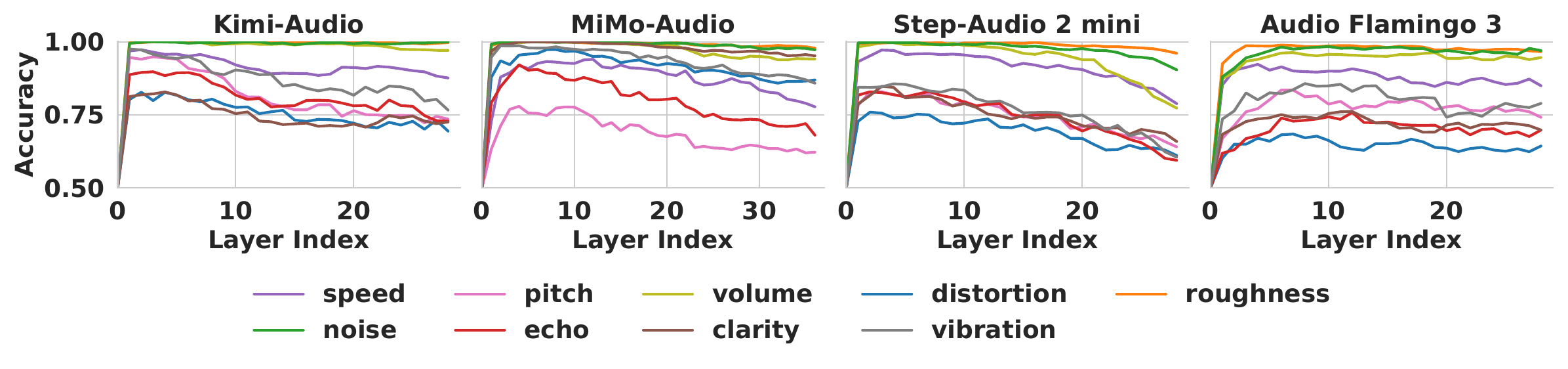}
\caption{Linear probing accuracies with text prediction representations across language decoder layers.}
\label{fig:linear-probe-text}
\end{figure}

%% file: tables/fine_tune_accuracy_table.tex
\begin{table}[t]
\centering
\small
\caption{Kimi-Audio fine-tuning accuracy (\%) by category.}
\label{tab:fine_tune_accuracy}
\begin{tabular}{lrrrrrrrrr}
\toprule
Method & speed & pitch & volume & distortion & roughness & noise & echo & clarity & vibration \\
\midrule
LM-Head & 73.2 & 57.0 & 88.2 & 52.2 & 97.7 & 98.6 & 61.5 & 57.4 & 53.3 \\
All-Linear & 99.0 & 98.7 & 100.0 & 88.5 & 100.0 & 100.0 & 96.8 & 87.0 & 99.2 \\
\bottomrule
\end{tabular}
\end{table}

%% file: figures/fine_tuning_visualization.tex
\begin{figure}[t]
\centering
\begin{minipage}[t]{0.38\linewidth}
    \centering
    \includegraphics[width=\linewidth]{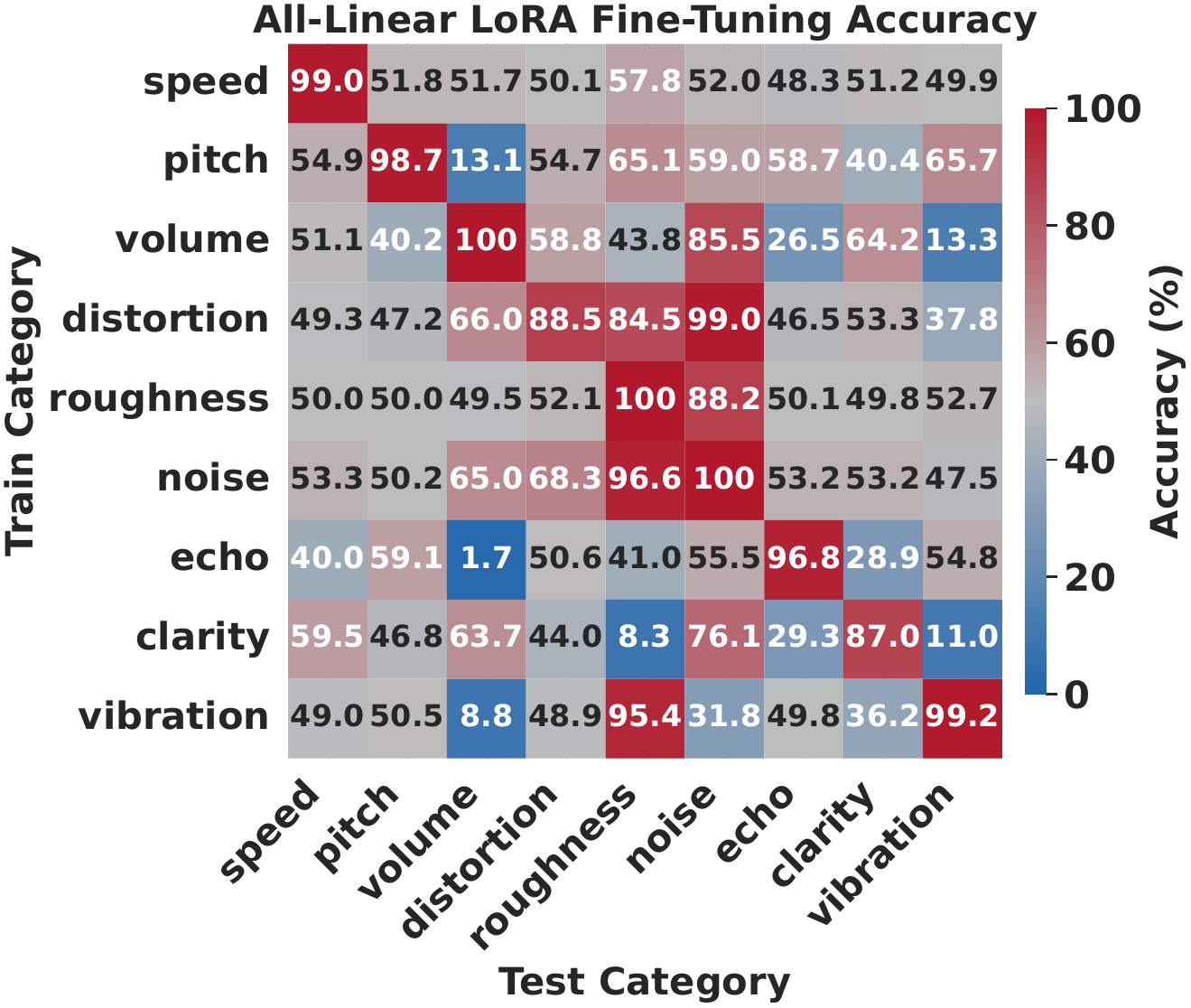}
    \vspace{-0.5em}
    \subcaption{Cross-category accuracies (\%) of all-linear LoRA fine-tuned model.}
    \label{fig:kimi-ft-heatmap}
\end{minipage}
\hfill
\begin{minipage}[t]{0.58\linewidth}
    \centering
    \includegraphics[width=\linewidth]{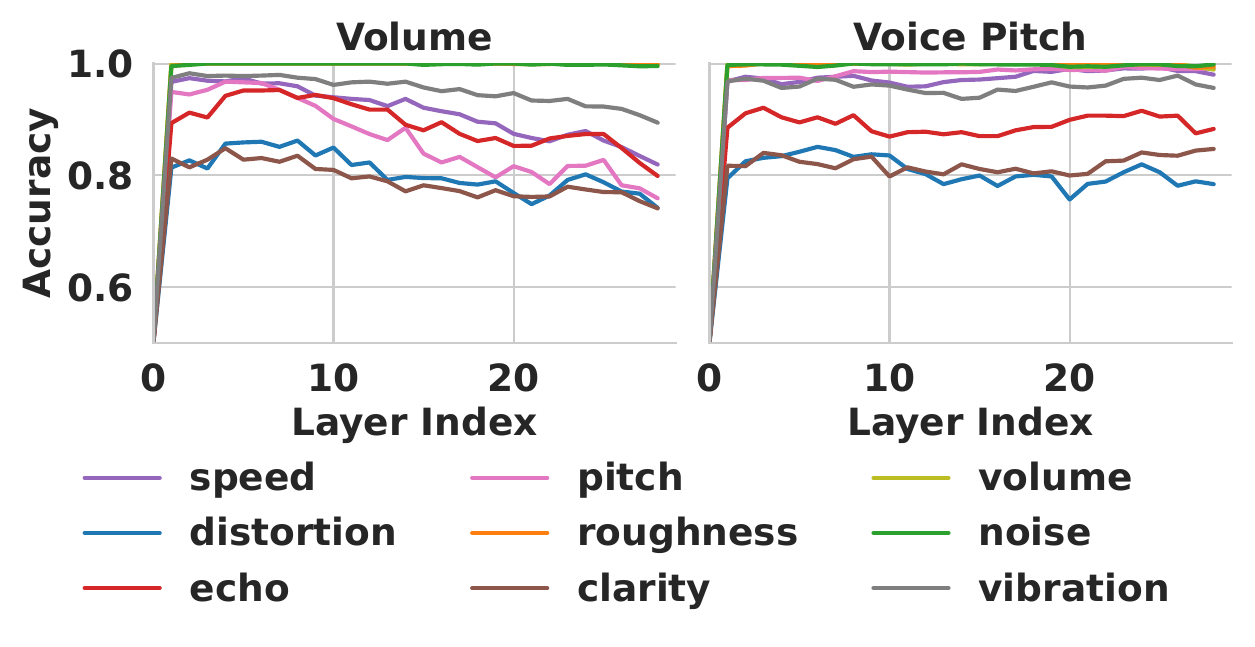}
    \vspace{-0.5em}
    \subcaption{Layer-wise linear probing accuracy after fine-tuning.}
    \label{fig:kimi-ft-probing}
\end{minipage}

\caption{
Analysis of LoRA fine-tuning effects on Kimi-Audio.
(Left) Cross-category performance.
(Right) Layer-wise probing accuracy for models fine-tuned on Volume and Voice Pitch.
}
\label{fig:kimi-ft-combined}
\end{figure}

%% file: chapters/6_conclusion.tex
\section{Conclusion}
\label{sec:conclusion}

In this paper, we introduced VocalGrad, a benchmark designed to evaluate the acoustic perception abilities of LALMs.
Using both text-based evaluation and linear probing, we systematically analyzed the behavior of state-of-the-art LALMs.
\par

Our results show that even simple acoustic changes that humans easily perceive remain challenging for current LALMs when framed as a text-based classification task.
Further analysis reveals two distinct failure modes.
For categories characterized by short-term energy variations, such as Volume, task-relevant information is preserved in internal representations up to the final layers, while the language model head acts as the primary bottleneck.
For other categories, such as Voice Pitch and Speaking Speed, the relevant information is initially extracted in early layers but becomes progressively less linearly decodable as it propagates through deeper layers.
\par

We further validate these findings through LoRA fine-tuning experiments.
For categories with preserved representations, performance can be substantially improved by updating only the language model head.
When all linear layers are fine-tuned, we observe not only improved task performance but also cross-category effects and enhanced linear separability in internal representations.
\par
As future work, we plan to investigate how improvements in acoustic perception can lead to consistent and generalizable performance gains on higher-level tasks such as emotion recognition and speech event understanding.
We hope that VocalGrad and the accompanying analysis provide useful insights toward building LALMs with more robust acoustic understanding capabilities.

%% file: chapters/appendix.tex
\section{Limitations}
\label{sec:limitations}

Although VocalGrad and our subsequent analyses reveal shortcomings in LALMs' acoustic perception abilities that are not captured by existing benchmarks, our work itself has several limitations.
\par

First, although VocalGrad reveals that current LALMs struggle to map isolated temporal acoustic changes to explicit direction judgments under controlled conditions, it does not establish that these findings directly generalize to naturally occurring acoustic variation.
In natural recordings, multiple acoustic factors may vary simultaneously and co-vary with one another, whereas VocalGrad intentionally manipulates one factor at a time.
Evaluating LALMs on naturally occurring, co-varying acoustic changes would therefore provide a complementary perspective and is an important direction for future work.
\par
Second, in the main linear probing experiments, we aggregate audio representations using average pooling over 16 fixed temporal segments.
Although the position-wise analysis in Appendix~\ref{sec:position-wise-probing} provides additional evidence that task-relevant information is distributed non-uniformly across token positions,
we do not systematically examine different temporal resolutions or aggregation strategies.
Information that is not recoverable under our probing setup may therefore remain accessible under finer-grained or nonlinear analyses.
A systematic comparison across different time resolutions, as well as the use of models better suited for temporal sequence modeling (e.g., LSTM~\citep{lstm}), would provide a more complete understanding of information retention in LALMs.
\par

Finally, we do not provide a detailed analysis of the differences between categories for which temporal changes remain decodable in deeper layers (e.g., Volume, Audio Roughness, Background Noise) and those for which they do not.
We hypothesize that this distinction may be related to the temporal characteristics of the relevant acoustic cues, such as whether they can be estimated from short temporal windows or require integration over longer contexts, but this hypothesis is not thoroughly validated.
Further analysis using established acoustic feature sets such as GeMAPS~\citep{gemaps} would be necessary to provide a clearer explanation of these differences.

\section{Detailed Experimental Settings}
\subsection{Acoustic Modeling in LALMs}
\label{sec:acoustic-modeling}
In this section, we explain how the four LALMs used in our experiments incorporate various design choices in their audio tokenizers, encoders, and training objectives to enhance their ability to process acoustic information.
\par
Audio~Flamingo~3~\citep{audioflamingo} introduces AF-Whisper, a unified audio encoder trained across speech, sound, and music.
The encoder is initialized from a pretrained Whisper large-v3 model and further trained using an audio captioning objective with next-token prediction.
The captions are generated by integrating multiple sources of information, including transcripts, ambient sound descriptions, and music-related metadata, enabling the model to learn a unified representation of diverse audio signals.
\par

Similarly, Step-Audio~2~\citep{stepaudio2} pretrains its audio encoder on a variety of speech and audio understanding tasks, including ASR, speaker attribute estimation, and audio event detection, to capture diverse acoustic information.
Notably, the pretrained audio encoder is kept frozen during subsequent training, while an adaptor is used to project its representations into the language model.
\par

Kimi-Audio~\citep{kimi-audio} adopts a hybrid audio representation consisting of discrete semantic tokens and continuous acoustic features.
The discrete tokens, derived from ASR-based tokenization, serve as the primary representation of audio, while continuous features extracted from a Whisper-based encoder are incorporated to enrich acoustic perception.
These representations are processed by a unified audio LLM with shared transformer layers and separate heads for text and audio token generation.
The model is trained with a combination of objectives, including unimodal modeling, audio-text mapping, and interleaved audio-text prediction, which emphasize alignment between audio and textual representations.

\par
MiMo-Audio~\citep{mimo-audio} adopts a fundamentally different approach by directly modeling high-fidelity acoustic tokens with an autoregressive objective.
It introduces a dedicated tokenizer trained from scratch using both reconstruction and audio-to-text objectives, aiming to preserve both semantic and acoustic information.
The resulting audio tokens are grouped into patches and processed by a large language model, which performs next-token or next-patch prediction over interleaved text and audio sequences.

This design allows the model to maintain a more complete representation of the original audio signal compared to approaches relying primarily on semantic tokens or text-aligned supervision.
\par

Despite these diverse approaches, it remains unclear whether they can reliably capture primitive acoustic variations.

\subsection{Detailed Description of Temporal Trajectories for Data Augmentation}
\label{sec:temporal-trajectory-details}

In this section, we provide a detailed description of the temporal trajectories
used for data augmentation in Section~\ref{sec:dataset-common-pipeline}.

Let $t\in[0,1]$ denote normalized time within the detected speech span, and let
$a_{\min}$ and $a_{\max}$ denote the minimum and maximum augmentation
strengths, respectively.
We define $\Delta a = a_{\max}-a_{\min}$.
For each trajectory, we define a trajectory-specific progress function $g(t)$.
Based on $g(t)$, the augmentation strengths for increasing and decreasing clips
are defined as
\begin{align}
a_{\mathrm{inc}}(t) &= a_{\min} + \Delta a\, g(t), \\
a_{\mathrm{dec}}(t) &= a_{\max} - \Delta a\, g(t).
\end{align}
We use the following four progress functions:
\begin{align}
\texttt{jump}: \quad
& g(t) =
\begin{cases}
0, & t < \frac{1}{2},\\
1, & t \ge \frac{1}{2},
\end{cases}
\\
\texttt{linear}: \quad
& g(t) = t,
\\
\texttt{quad\_first\_flat}: \quad
& g(t) = t^2,
\\
\texttt{quad\_last\_flat}: \quad
& g(t) = 1-(1-t)^2.
\end{align}
Accordingly, \texttt{quad\_first\_flat} has zero slope at $t=0$, whereas
\texttt{quad\_last\_flat} has zero slope at $t=1$.

The transformations are applied at discrete positions in each input clip.
For the Echo and Voice Vibration categories, the augmentation strength is
updated at each audio sample, whereas for the other seven categories, it is
updated in quantized 50-ms windows.

For the Audio Distortion and Audio Roughness categories, stronger augmentation
corresponds to a lower effective sampling rate and bit depth, respectively.
Therefore, the mapping between $a_{\mathrm{inc}}(t)$ and $a_{\mathrm{dec}}(t)$
and the increasing/decreasing labels are reversed for these two categories.

\subsection{Hyperparameters for Dataset Generation}
\label{sec:dataset-hyperparameters}

In this section, we list the concrete hyperparameters used to generate category-specific augmentations described in Section~\ref{sec:transformation-categories}.

\paragraph{Speaking Speed (\textit{speed})}
We use librosa~\citep{librosa} to modify the speaking rate of each speech segment.
Given a hyperparameter $\Delta_{\max}$, the time-stretching factor is varied from $1 - \Delta_{\max}$ to $1 + \Delta_{\max}$ (or vice versa) over the speech segment.
We use $\Delta_{\max} = (0.4, 0.3, 0.2)$ for the high, medium, and low intensity tiers, respectively.

\paragraph{Voice Pitch (\textit{pitch})}
We use librosa~\citep{librosa} to shift the pitch of each speech segment.
The pitch is gradually shifted from $-\Delta_{\max}$ to $+\Delta_{\max}$ semitones (or vice versa) across the speech segment.
We use $\Delta_{\max} = (3, 2, 1)$ for the high, medium, and low intensity tiers, respectively.

\paragraph{Volume (\textit{volume})}
We apply gain control to adjust the loudness of each speech segment.
The gain is varied from $-\Delta_{\max}$ dB to $+\Delta_{\max}$ dB (or vice versa) over time.
We use $\Delta_{\max} = (18, 12, 6)$ for the high, medium, and low intensity tiers, respectively.

\paragraph{Audio Distortion (\textit{distortion})}
We simulate distortion caused by low-resolution audio~\citep{distortion-literature}.
Each clip is downsampled from the original sampling rate (48 kHz) to $F_{\min}$ (or vice versa), without applying anti-aliasing filtering.
To emphasize distortion artifacts, we hold constant sample values within each downsampled interval.
We use $F_{\min} = (6, 12, 18)$ kHz for the high, medium, and low intensity tiers, respectively.

\paragraph{Audio Roughness (\textit{roughness})}
We simulate quantization artifacts by reducing bit depth.
The waveform is quantized from the default 16-bit depth down to $b_{\min}$ bits, producing a ``bitcrusher'' effect~\citep{bitcrusher} associated with perceptual roughness~\citep{psychoacoustics}.
Although floating-point values are assigned during interpolation, the bit depth is rounded to integer values during augmentation.
We use $b_{\min} = (4, 6, 8)$ for the high, medium, and low intensity tiers, respectively.

\paragraph{Background Noise (\textit{noise})}
We generate additive Gaussian noise and add it to each clip.
Let $R$ denote the mean RMS of the detected speech segment.
The noise level is controlled such that its RMS varies from $0$ to $R \times r_{\max}$ (or vice versa).
We use $r_{\max} = (0.12, 0.07, 0.03)$ for the high, medium, and low intensity tiers, respectively.

\paragraph{Echo (\textit{echo})}
We apply a reverberation-based echo effect using a Schroeder reverberator~\citep{reverbation}.
A fully reverberated signal is generated and mixed with the original signal using a time-varying mixing coefficient.
The augmentation strength is controlled by the reverberation spread and the mixing coefficient.
We use $s_{\max} = (0.9, 0.75, 0.55)$ for the maximum reverberation spread across the high, medium, and low tiers.
The mixing coefficient ranges from 0 to $c_{\max}$, where $c_{\max} = (0.52, 0.40, 0.28)$ for the high, medium, and low tiers, respectively.

\paragraph{Voice Clarity (\textit{clarity})}
We simulate reduced speech clarity by applying low-pass filtering in the frequency domain.
Using librosa~\citep{librosa}, we perform STFT and ISTFT to remove high-frequency components.
To avoid unintended loudness changes, we normalize the RMS of the filtered signal.
The cutoff frequency is varied from half the sampling rate down to $F_{\min}$ (or vice versa).
We use $F_{\min} = (1.2, 2.5, 5.0)$ kHz for the high, medium, and low intensity tiers, respectively.

\paragraph{Voice Vibration (\textit{vibration})}
We apply low-frequency amplitude modulation to simulate vocal vibration effects.
The modulation frequency varies from 0 Hz to 20 Hz, and the modulation depth varies from 0 to $A_{\max}$.
In the decreasing condition, both parameters are reversed.
We use $A_{\max} = (0.9, 0.75, 0.6)$ for the high, medium, and low intensity tiers, respectively.

\par
If any augmentation produces values outside the valid pulse-code modulation range, the waveform is clipped to the allowable range.
For the Volume category, approximately 0.4\% of samples are clipped.
For all other categories, fewer than $10^{-6}$ of samples are affected.

\subsection{Human Annotation}
\label{sec:human-annotation-tool}
\input{figures/human-annotation-ui}

Human performance on VocalGrad reported in Section~\ref{sec:benchmarking-results} was collected using the following procedure.
We recruited six annotators, all of whom are regularly exposed to English in daily communication.
\par

Each participant was assigned up to five categories and asked to answer 50 binary questions per category.
For each category, the 50 test clips contain equal numbers of increasing and decreasing examples, with approximately balanced coverage of the three intensity tiers and four temporal trajectories.
However, an implementation error in the sampling procedure coupled direction with trajectory: \texttt{linear} and \texttt{quad\_last\_flat} were sampled only for the increasing direction, whereas \texttt{jump} and \texttt{quad\_first\_flat} were sampled only for the decreasing direction.
Consequently, the reported human accuracies describe performance on this sampled subset, and the trajectory-wise results do not isolate the effect of trajectory from that of direction.
\par

The annotation was conducted using a dedicated web-based tool, shown in Figure~\ref{fig:human-annotation-ui}.
Participants were provided with a PDF manual explaining how to use the tool.
They completed the task in their preferred environment.
The interface displays the current progress, the prompt associated with the audio clip, a playback button, two buttons for assigning labels (``increase'' and ``decrease''), and a submission button.
Since some annotators are non-native English speakers, we allowed them to consult external resources to understand the attribute terms if necessary.
We also allowed multiple replays of each audio clip before submitting an answer.
\par

All participants were informed about the purpose of the study and that participation was voluntary.
We offered all participants compensation at a rate of 3,000 JPY per hour.
Since only binary labels were collected and only aggregated results are reported, the study poses minimal risk of personal data exposure.

\subsection{Runtime Configuration of Models}
\label{sec:model-runtime-parameter}

\paragraph{LALM Text Generation}
All experiments with open-source LALMs are conducted using the Transformers library~\citep{transformers}.
When necessary, we additionally use model-specific runtime libraries provided by the authors~\footnote{https://github.com/MoonshotAI/Kimi-Audio}~\footnote{https://github.com/XiaomiMiMo/MiMo-Audio}.
\par

Unless otherwise specified, models are run with \texttt{bfloat16} precision.
For Audio~Flamingo~3, the model is executed with the default \texttt{float32} precision.
\par

For MiMo-Audio, we use \texttt{temperature}=0.3 and \texttt{top\_p}=0.95.
To ensure reproducibility, we fix the random seed before each generation.
For the other open-source models, we use deterministic greedy decoding when available.
\par

For Gemini 3 Flash, the \texttt{thinking\_level} parameter is set to ``low'' to reduce the effect of reasoning and better isolate perceptual capability.
Following the official API recommendation~\footnote{https://ai.google.dev/gemini-api/docs/gemini-3}, we use non-deterministic sampling with \texttt{temperature}=1.0.
\par

By default, we limit the maximum number of generated tokens per input to 128.
If the output cannot be parsed on the first attempt, we rerun the model with a maximum of 256 tokens to reduce parsing errors.

\paragraph{Linear Classifiers}
Linear classifiers for probing experiments are implemented in PyTorch~\citep{pytorch}.
Given an input feature dimension $d$, we use a \texttt{Linear} layer of shape $(d, 1)$~\footnote{https://docs.pytorch.org/docs/main/generated/torch.nn.Linear.html}, trained with \texttt{BCEWithLogitsLoss}~\footnote{https://docs.pytorch.org/docs/2.11/generated/torch.nn.BCEWithLogitsLoss.html}.
\par

We use AdamW~\citep{adamw} as the optimizer with learning rate $10^{-3}$, weight decay $10^{-4}$, batch size $512$, and $200$ epochs.

\paragraph{LoRA Fine-tuning}
We perform LoRA~\citep{lora} fine-tuning using the PEFT framework~\citep{peft}.
The LoRA hyperparameters are set to $\alpha=16$, rank $r=8$, and dropout $0.05$, following common practice.
\par

To ensure fair comparison between LM-Head and All-Linear settings, we limit training to one epoch in both cases, as All-Linear training is significantly more computationally expensive.
From the 12,000 training samples per category, we reserve 10\% for validation, resulting in 10,800 training samples.
\par

We use AdamW~\citep{adamw} with learning rate $2\times10^{-4}$ and weight decay $0$.
The batch size is set to $1$, with gradient accumulation over 16 steps.

\section{Additional Experimental Results}
\subsection{Per-Temporal-Trajectory Analysis of Benchmarking Results}
\label{sec:temporal-trajectory-analysis}
\input{tables/trajectory-performance}

Table~\ref{tab:trajectory-performance} provides a per-temporal-trajectory breakdown of the human and model performance on VocalGrad presented in Section~\ref{sec:benchmarking-results}.
Due to the sampling implementation error described in
Appendix~\ref{sec:human-annotation-tool}, trajectory and ground-truth direction are coupled in the human-evaluation subset.
Therefore, human results do not distinguish the effects of trajectory type from those of ground-truth direction.
\par

Human accuracy was highest for the jump trajectory within this subset, while all models remained near chance across all trajectory types.
Therefore, we do not observe a clear systematic effect of trajectory type on model performance under the current near-chance performance levels.

\subsection{Per-Intensity-Tier Analysis of Benchmarking Results}
\label{sec:intensity-tier-analysis}

\input{figures/background_noise_voice_pitch_pair}
Some of the errors made by human annotators in Section~\ref{sec:benchmarking} can be explained by the intensity tiers introduced in VocalGrad.
Figure~\ref{fig:bgnoise-pitch-tier-accuracy} shows tier-wise accuracy for the Background Noise and Voice Pitch categories.

For the Voice Pitch category, human performance decreases as the transformation intensity becomes weaker, indicating that lower-intensity changes are perceptually more difficult.
This confirms that the intensity tiers successfully control task difficulty and enable more fine-grained analysis.

For the Background Noise category, human annotators achieve near-perfect accuracy across all tiers.
In contrast, LALMs show a modest improvement in performance for higher-intensity transformations.
This suggests that increasing transformation strength can partially mitigate the difficulty for models, although performance remains close to chance overall.

\subsection{Analysis of Parsing Failure in Benchmarking}
\label{sec:parsing-failure-analysis}

\input{tables/parsing_output_match}
\input{tables/exact-match-accuracy}
\input{tables/category_non_exact_match}

As described in Section~\ref{sec:evaluation-settings}, we map model outputs to direction labels using case-insensitive, whole-word matching of “increase”/“increases” and “decrease”/“decreases”.
An output is assigned a label only if it contains exactly one of the two directions; otherwise, it is counted as incorrect. Accuracy is computed over all evaluated clips.
\par

We adopt this strategy due to the non-negligible parsing errors observed under the exact-match condition.
Table~\ref{tab:vocalgrad-parsing-output-match} shows the proportion of outputs that match the target labels under both exact-match and rule-based conditions.
While the parse error rate is negligible for all open-source models, Gemini 3 Flash achieves only 83.9\% exact-match rate.
This increases to 99.1\% under rule-based parsing.
To ensure fair comparison across models, we therefore use rule-based parsing for evaluation.
\par

The accuracies computed under the exact-match condition are reported in Table~\ref{tab:vocalgrad-exact-match-accuracy}.
While the performance of open-source models remains largely unchanged, Gemini 3 Flash shows a significant drop in accuracy across several categories.
\par

Qualitative analysis of Gemini 3 Flash outputs reveals three common patterns of non-exact-match responses:
\begin{itemize}
    \item Full-sentence responses that clearly indicate one of the options (e.g., \textit{The audio roughness increases over time.})
    \item Incomplete responses, likely truncated due to token limits (e.g., \textit{The audio roughness in the provided clip does not})
    \item Refusal or uncertainty responses (e.g., \textit{The provided audio clip is too short to determine if the distortion increases or decreases over time.})
\end{itemize}
Rule-based parsing correctly recovers the intended label in the first case.
\par

The frequency of non-exact-match outputs also varies significantly across categories.
Table~\ref{tab:vocalgrad-category-non-exact-match} shows the per-category breakdown.
While the average non-exact-match rate for Gemini 3 Flash is 16.1\%, it reaches 67.1\% for the Audio Roughness category.
This suggests that the likelihood of instruction-following behavior depends on the acoustic properties of the input or the specific attribute queried in the prompt.

\subsection{Statistical Significance of the Benchmarking Results}
\label{sec:statistical-significance}

\input{figures/human_model_accuracy_ci_grid}

We assess the statistical significance of the results reported in Section~\ref{sec:benchmarking-results} using binomial simulation-based confidence intervals.
For each VocalGrad category and each LALM (or the group of human annotators), we simulate prediction outcomes by sampling from a binomial distribution parameterized by the observed accuracy.
We draw samples matching the original number of predictions ($n = 150$ for human annotators and $n = 2400$ for LALMs), repeat this process 10{,}000 times, and report the 2.5 and 97.5 percentiles as the 95\% confidence interval.
\par

The results are shown in Figure~\ref{fig:human-model-accuracy-ci-grid}.
Most model confidence intervals overlap with the 50\% baseline, indicating that their performance is statistically indistinguishable from chance.
Despite the relatively smaller sample size for human annotations, the confidence intervals clearly demonstrate a substantial performance gap between humans and models.
\par

Interestingly, we also observe cases where model accuracy falls significantly below 50\%, such as MiMo-Audio on the Audio Roughness category.
Since parsing errors (Appendix~\ref{sec:parsing-failure-analysis}) do not significantly affect this category, this result suggests that the model may partially perceive the acoustic feature but fails to consistently map it to the correct label.

\subsection{Failure Analysis with Output Distribution}
\label{sec:benchmark-failure-analysis}
\input{tables/original_vs_source_language_bias}

Some of the failure modes in Section~\ref{sec:benchmarking} can be explained by examining the output distribution of each LALM.
Table~\ref{tab:original-source-increase-response} reports the proportion of ``increase'' responses for each model on VocalGrad over successfully parsed responses.
For reference, we also apply the same binary prompts to the original VCTK source clips and measure the corresponding response distribution.
\par

The results reveal that some models exhibit strong directional bias in their outputs.
In particular, Step-Audio~2~mini and Audio~Flamingo~3 show extreme bias toward a single response across categories, which provides a straightforward explanation for their near-chance performance.
\par

Even for models that do not exhibit strong bias when aggregated across categories, significant biases emerge at the category level.
For example, Gemini 3 Flash predicts ``increase'' for only 2.1\% of the Echo samples, whereas MiMo-Audio predicts ``increase'' for 90.9\% of the same category.
Furthermore, when evaluated on the original VCTK source clips, these biases become even more pronounced.
\par

These observations suggest that response bias plays a significant role in model performance under the binary text-answering format.
In particular, when models lack reliable acoustic evidence, they tend to default to biased or degenerate response patterns.
\par

One possible explanation is that, in the absence of sufficient signal, the language model head falls back on prior biases or unstable decision boundaries, as observed in prior work on LLM output behavior~\citep{stochasticity}.
Consistent with this hypothesis, the linear probing analysis in Section~\ref{sec:linear_probing} shows that for certain transformations, such as audio roughness, background noise, and volume, the relevant information is preserved in the hidden representations but not effectively utilized by the language model head.

\subsection{Prompt Sensitivity of Benchmarking Results}
\label{sec:prompt-sensitivity}

We analyze the sensitivity of benchmarking results reported in Section~\ref{sec:benchmarking-results} to variations in prompt design.

\paragraph{Prompt Variants}
\input{tables/human_model_accuracy_comparison_swap_order}
\input{tables/human_model_accuracy_comparison_audio_ref}

We replace the base prompt with the following two variants:
\begin{itemize}
    \item \texttt{swap\_order}: Swap the order of labels in the prompt. \textit{Does the \{attribute\} decrease or increase over time? Answer with only one word: ``decrease'' or ``increase''.}
    \item \texttt{audio\_ref}: Explicitly emphasize that the question refers to the input audio. \textit{Does the \{attribute\} of this audio clip increase or decrease over time? Answer with only one word: ``increase'' or ``decrease''.}
\end{itemize}
\par

Table~\ref{tab:vocalgrad-human-model-accuracy-swap-order} and Table~\ref{tab:vocalgrad-human-model-accuracy-audio-ref} report the results for open-source models under the \texttt{swap\_order} and \texttt{audio\_ref} prompts, respectively.
Across all models and prompt variants, performance remains close to chance level, with no meaningful improvement over the base prompt.
These results suggest that simple variations in prompt phrasing do not resolve the observed performance limitations.

\paragraph{Interchanged Attribute Words}
\input{figures/vocalgrad-cross-attribute}

We further analyze how the choice of attribute words affects benchmarking performance of LALMs on VocalGrad.
Specifically, we replace the query attribute word with those corresponding to other categories and measure the resulting accuracy on Kimi-Audio and MiMo-Audio.
\par

The results are shown in Figure~\ref{fig:vocalgrad-cross-attribute-kimi-mimo}.
Notably, for certain categories such as Audio Roughness and Background Noise, Kimi-Audio achieves performance significantly above or below chance depending on the query word.

As shown in Section~\ref{sec:linear_probing_results}, internal representations of LALMs preserve sufficient information for these categories up to late layers.
The cross-attribute results suggest that performance differences arise not from the absence of acoustic information, but from the model’s ability to associate the perceived signal with the corresponding attribute word.

\par

A limitation of this analysis is that for some categories, such as Speaking Speed and Voice Pitch, there are no alternative attribute words within the VocalGrad categories that carry similar meanings.
We therefore conduct a more direct evaluation using category-specific query
rewriting below.

\paragraph{Category-specific Query Rewriting}
\input{tables/query-patterns}
\input{tables/alternative-query-performance}
While the preceding two experiments use prompt modifications that can be
applied consistently across augmentation categories, we additionally construct
category-specific query variants for selected categories, allowing larger
changes in both lexical choice and query structure.
\par

Table~\ref{tab:alternative-query-patterns} lists the query variants used in
this experiment.
Each variant uses different attribute wording and sentence structure from the
base prompt used in Section~\ref{sec:benchmarking}.
\par

We evaluate Kimi-Audio on the corresponding VocalGrad clips using each prompt.
Semantically increasing responses, such as ``rise,'' ``upward,'' ``faster,''
and ``accelerate,'' are mapped to the increase label, while their opposites
are mapped to the decrease label.
\par

Table~\ref{tab:alternative-query-performance} reports classification accuracy
and the proportion of successfully parsed outputs mapped to the increasing direction.
None of the alternative prompts substantially improves accuracy, which remains
between 50.0\% and 53.2\% across all nine prompts.
However, prompt wording strongly affects the output distribution.
For example, the proportion of increase predictions for Voice Pitch changes
from 0.1\% with ``tone rise or fall'' to 40.1\% with
``frequency rise or decline,'' while accuracy remains close to chance.
Similarly, for Speaking Speed, the proportion increases from 0.9\% to
26.4\% across prompt variants, while the highest accuracy is only 53.2\%.
\par

These results support two conclusions.
First, text-output behavior is sensitive to query phrasing.
Second, changing the surface wording can substantially shift the model's
answer bias without reliably improving its sensitivity to the actual direction of the acoustic transformation.
Thus, for Voice Pitch and Speaking Speed, the near-chance performance cannot be explained solely by a poorly chosen attribute word or answer phrasing.

\subsection{Few-shot Prompting}
\input{tables/vocalgrad-fewshot}

For the three open-source models that support interleaved audio-text inputs, we evaluate whether few-shot prompting improves benchmarking performance.
Specifically, for each query in VocalGrad, we randomly sample four examples from the training set within the same category, two with ``increase'' labels and two with ``decrease'' labels.
These examples are randomly ordered and prepended to the input in the following format:
\textit{Example \{i\}: \{audio\} $\rightarrow$ \{increase/decrease\}}.
\par

The results are reported in Table~\ref{tab:vocalgrad-fewshot}.
While a slight improvement is observed for Step-Audio~2~mini, overall performance remains close to chance level, indicating that few-shot prompting does not substantially improve model performance on this task.

\subsection{Benchmarking Results with Different Source Datasets}
\label{sec:different-sources}
\input{tables/source-domain-expansion}

As described in Section~\ref{sec:dataset_construction}, the main VocalGrad benchmark uses speech clips from VCTK as source clips.
To examine whether the observed performance limitations persist across more diverse source domains, we apply the same VocalGrad data augmentation and benchmarking pipeline to speech clips from the following two additional datasets:
\begin{itemize}
    \item Loquacious Set~\citep{loquacious}, which contains diverse English speech recorded under both clean and noisy conditions.
    \item Common Voice~\citep{commonvoice}, which contains multilingual speech clips including both spontaneous and scripted speech.
    To obtain a source distribution distinct from the read English speech in VCTK, we specifically use spontaneous-speech subsets from five low-resource languages: Kabardian, Puno Quechua, Tashlhiyt, Eastern Min, and Chiga.
\end{itemize}

For each dataset, we select 100 source clips to match the number of source clips used in the original VCTK-based benchmark.
For Loquacious Set, the selected clips are balanced across male and female speakers.
For Common Voice, the clips are balanced across the five target languages.
We additionally limit the duration of each selected clip to at most 30 seconds.

Table~\ref{tab:source-domain-expansion} reports the performance of Kimi-Audio across the three source datasets, including the original VCTK results.
Performance remains close to chance overall, with average accuracies of 50.6\% on Loquacious Set and 52.9\% on Common Voice.
Although some category--dataset combinations, such as Speaking Speed on Common Voice, show moderately higher accuracy, the overall failure pattern is consistent with that observed on the VCTK-based benchmark.

\subsection{Benchmarking Results with Different Problem Formats}
\label{sec:different-problem-formats}
\input{tables/different-problem-performance}
\input{tables/different-problem-prediction-distribution}

As described in Section~\ref{sec:prompting-format}, the main VocalGrad benchmark is formulated as a binary classification task that predicts the direction of temporal acoustic change.
However, by combining the augmented clips with the original source clips, VocalGrad can also be used to evaluate LALMs under alternative problem formats.
In this section, we benchmark Kimi-Audio using two such formats and analyze its behavior.
\par

Specifically, we consider the following two additional problem formats:
\begin{itemize}
    \item \textbf{Change Detection} (Yes/No):
    We use the prompt ``Does the \{attribute\} vary with time?'' to formulate a binary task that detects the presence of a controlled temporal change.
    The unmodified source clips are used as negative examples, while the VocalGrad-augmented clips are used as positive examples.

    \item \textbf{Ternary Question} (Ternary):
    We use the prompt ``How does the \{attribute\} vary with time? Answer with only one word: `increase', `decrease', or `constant'.''
    Increasing and decreasing VocalGrad clips are assigned their corresponding labels, while the unmodified source clips are assigned the ``constant'' label.
\end{itemize}
\par

Table~\ref{tab:alternative-problem-format-performance} reports the performance of Kimi-Audio under these alternative problem formats.
Kimi-Audio does not achieve reliable performance under either format.
In the Yes/No task, macro-F1 remains close to chance across categories.
In the ternary task, macro-F1 ranges from 21.6\% to 27.6\%, below the 33.3\% chance level for balanced three-class classification.
Thus, reformulating the original increase/decrease task as either binary change detection or ternary classification does not resolve the observed end-to-end failure.
\par

We also examine the prediction distributions separately for increasing, decreasing, and source clips.
Table~\ref{tab:alternative-problem-format-distribution} reports the resulting distributions.
Consistent with the output-bias analysis in Appendix~\ref{sec:benchmark-failure-analysis}, Kimi-Audio exhibits a strong format-dependent preference for a single answer: ``yes'' in the Yes/No task and ``decrease'' in the ternary task.
More importantly, the prediction distributions differ only modestly across increasing, decreasing, and source clips.
For example, the model predicts ``decrease'' for approximately 89--92\% of all three clip types in the ternary setting.
This suggests that the observed failure is not specific to the original increase/decrease binary formulation and that similar answer-dominance behavior occurs under the two additional problem formats.
\par

One limitation of this experiment is that the unmodified source speech is not necessarily acoustically constant.
Natural speech contains local temporal fluctuations in attributes such as pitch, volume, and speaking speed.
We nevertheless use the source clips as the ``constant'' class because they contain no systematically imposed VocalGrad trajectory.
Therefore, the ``constant'' label should be understood as indicating the absence of a controlled augmentation rather than strict physical constancy throughout the clip.
This label approximation may introduce ambiguity, particularly for speaker-intrinsic attributes.
\par

However, this ambiguity alone is unlikely to explain the observed prediction pattern, since Kimi-Audio's overprediction of ``decrease'' occurs not only for source clips but also for clips with explicitly increasing and decreasing transformations.
Therefore, these additional experiments primarily demonstrate that the model's strong output bias and poor discrimination among clip types persist across alternative task formulations.

\subsection{Linear Probing Across Different Token Positions}
\label{sec:position-wise-probing}
\input{tables/position-probing-query-audio}
\input{tables/position-probing-audio-query}

The probing analysis in Section~\ref{sec:linear_probing_results} uses
temporally aggregated audio representations and the representation at the
output-prediction position.
While this analysis reveals how task-relevant information changes across
model stages and decoder layers, it does not show how such information is
distributed across token positions.
To investigate this aspect, we conduct position-wise linear probing on
Kimi-Audio.
\par

We first divide the audio-token sequence into four equal temporal segments
and probe the representation at the final token of each segment.
Together with the output-prediction position, we consider the following
five positions:
\begin{itemize}
    \item $a_i$, $i \in \{1,\ldots,4\}$ (\textbf{Audio Segment End}):
    The final audio token of the $i$-th temporal segment.
    \item $o$ (\textbf{Output-Prediction Position}):
    The position used to predict the output token.
\end{itemize}
\par

We additionally examine how task-relevant information is represented at
text-query positions.
Under the default Query$\rightarrow$Audio ordering, query tokens precede the
audio tokens and therefore cannot attend to the subsequent audio tokens under
causal attention.
To enable analysis of audio-derived information at query positions, we
additionally evaluate an Audio$\rightarrow$Query ordering.
We verify that swapping the context order does not materially alter the
end-to-end benchmark performance of Kimi-Audio, which remains near chance
(50.6\% average accuracy).
\par

Under the Audio$\rightarrow$Query ordering, following the position-based
analysis of \citet{audiolens}, we additionally probe the following three
text positions:
\begin{itemize}
    \item $q_{attr}$ (\textbf{Attribute Last Token}):
    The final token corresponding to the attribute expression in the query
    (e.g., ``volume'').
    \item $q_p$ (\textbf{Query Penultimate Token}):
    The penultimate token of the input query.
    \item $q_l$ (\textbf{Query Last Token}):
    The final token of the input query.
\end{itemize}
\par

The probing results under the Query$\rightarrow$Audio and
Audio$\rightarrow$Query orderings are presented in
Table~\ref{tab:position-probing-query-audio} and Table~\ref{tab:position-probing-audio-query}, respectively.
Two patterns emerge.
First, the representations at $a_2$ and $a_3$ consistently contain less
linearly decodable task-relevant information than those at $a_1$ and $a_4$
across the examined layers.
Second, under the Audio$\rightarrow$Query ordering, task-relevant information
at text positions becomes more concentrated at the final query and
output-prediction positions in deeper layers, rather than remaining localized
at the attribute position.
\par

These results show that task-relevant information is not uniformly distributed
across token positions and that its distribution changes as representations
propagate through the language decoder.
Further investigation is required to disentangle the effects of source-audio
content, temporal position, and model-specific processing.

\subsection{Fine-tuning Results in All Target Models}
\label{sec:fine-tuning-all}

\input{tables/finetune-all-results}

We present the results of LoRA fine-tuning on the four open-source models in Table~\ref{tab:finetune-all-results}.
For LM-Head fine-tuning, the performance trends are consistent with the linear probing results on the late LLM layers reported in Section~\ref{sec:linear_probing_results}.
For example, MiMo-Audio achieves relatively higher linear probing accuracy on the Voice Clarity category compared to other models, and this trend is reflected in the LM-Head fine-tuning results.
For All-Linear fine-tuning, all four models show substantial performance improvements across most categories.

\subsection{Representation Alignment Across Categories After Fine-Tuning}
\label{sec:fine-tuning-category-alignment}
\input{tables/fine-tuning-category-alignment}
\input{figures/fine-tuned-model-similarity-matrix}

Section~\ref{sec:fine-tuning} shows that fine-tuning an LALM on one VocalGrad
category can affect its performance on other categories.
Based on that observation, we further investigate whether this cross-category effect is accompanied by the emergence of shared representational directions across categories.
\par

Specifically, we characterize each category by an
\textit{increase--decrease direction} in the hidden representation space and
compare these directions across categories.
Let $A_{c,\mathrm{inc}}$ and $A_{c,\mathrm{dec}}$ denote the sets of
VocalGrad test clips in category $c$ with increasing and decreasing labels,
respectively.
Let $M_l(a)$ denote the hidden representation at the output-prediction
position in layer $l$ for audio clip $a$.
We define the normalized increase--decrease direction
$\mathbf{d}_{l,c}$ as
\begin{align}
\bar{\mathbf{h}}_{l,c}^{\mathrm{inc}}
&=
\frac{1}{|A_{c,\mathrm{inc}}|}
\sum_{a \in A_{c,\mathrm{inc}}} M_l(a), \\
\bar{\mathbf{h}}_{l,c}^{\mathrm{dec}}
&=
\frac{1}{|A_{c,\mathrm{dec}}|}
\sum_{a \in A_{c,\mathrm{dec}}} M_l(a), \\
\mathbf{d}_{l,c}
&=
\frac{
\bar{\mathbf{h}}_{l,c}^{\mathrm{inc}}
-
\bar{\mathbf{h}}_{l,c}^{\mathrm{dec}}
}{
\left\|
\bar{\mathbf{h}}_{l,c}^{\mathrm{inc}}
-
\bar{\mathbf{h}}_{l,c}^{\mathrm{dec}}
\right\|_2
}.
\end{align}
We then measure the alignment between categories $c$ and $c'$ using
the cosine similarity
$\mathbf{d}_{l,c}^{\top}\mathbf{d}_{l,c'}$.
\par

Table~\ref{tab:category-direction-cosine} reports the mean signed cosine
similarity and mean absolute cosine similarity across all pairs of categories
for the base and fine-tuned Kimi-Audio models.
While the mean signed similarity remains relatively close to zero, the mean
absolute similarity increases substantially after fine-tuning, particularly
in the middle-to-late layers.
For example, at layer 28, the mean absolute cosine similarity increases from
0.205 for the base model to 0.729 and 0.682 for the Volume-tuned and
Voice-Pitch-tuned models, respectively.
These results suggest that category-specific fine-tuning can induce stronger
alignment of increase--decrease directions across categories, including
categories not used for fine-tuning.
\par

Figure~\ref{fig:fine-tuned-model-similarity-matrix} shows the corresponding
per-category similarities at layer 28.
In the fine-tuned models, many category pairs exhibit cosine similarities
with large absolute magnitudes, with both positive and negative signs.
This indicates that their increase--decrease directions become approximately
parallel or antiparallel, while the orientation of the direction remains
category dependent.
The coexistence of positive and negative alignments also explains why the
mean signed cosine similarity remains close to zero despite the substantial
increase in absolute similarity.
\par

However, this alignment does not emerge uniformly across categories.
In particular, Speaking Speed remains relatively weakly aligned with several
other categories even after fine-tuning.
This is notable because the Voice-Pitch-tuned model shows improved linear
probing performance on Speaking Speed despite its comparatively weak
directional alignment.
Thus, the cross-category effects of VocalGrad fine-tuning cannot be explained
solely by the emergence of a single shared increase--decrease direction.
Fine-tuning may also alter category-specific representations or their linear
separability in ways that are not captured by this direction-based analysis.
Further investigation is required to disentangle these effects.

\subsection{Transfer of VocalGrad Fine-Tuning to Emotion Recognition}
\label{sec:emotion-recognition}
\input{tables/meld-generalization}

To examine whether the effects of VocalGrad-based fine-tuning extend beyond the controlled benchmark setting, we evaluate the Kimi-Audio models fine-tuned with VocalGrad clips in Section~\ref{sec:fine-tuning} on the MELD~\citep{meld} emotion recognition dataset.
\par

We specifically evaluate models fine-tuned on the Volume and Voice Pitch categories.
MELD contains natural conversational speech and defines seven emotion labels: anger, disgust, fear, joy, neutral, sadness, and surprise.
For each sample, we score the option-letter tokens corresponding to the seven emotion labels using the model’s output logits and select the highest-scoring option.
\par

Table~\ref{tab:meld-transfer} reports precision, recall, and F1 for each class, together with their macro averages.
Voice Pitch fine-tuning produces a substantial shift in the model's prediction behavior.
In particular, recall for Joy increases to 68.9\%, while performance on several other classes declines.
Among samples predicted as Neutral by the base model, 25.8\% are instead predicted as Joy after Voice Pitch fine-tuning.
This suggests that category-specific VocalGrad fine-tuning can alter decision tendencies on natural downstream speech, but does not necessarily improve overall emotion recognition and may distort class-specific prediction behavior.
\par

Neither VocalGrad fine-tuning condition improves the overall macro-F1 on MELD.
The base model achieves a macro-F1 of 41.3, compared with 37.7 after Volume fine-tuning and 34.2 after Voice Pitch fine-tuning.
Volume fine-tuning slightly improves the F1 score for Neutral from 66.4 to 67.3, but this class-specific improvement does not translate into an overall gain.
\par

These results suggest that further study is needed to establish a direct connection between improvements on VocalGrad and downstream task performance.
The effects of VocalGrad fine-tuning are not entirely confined to the synthetic benchmark distribution, as the fine-tuned models exhibit altered prediction behavior on natural conversational speech.
However, the current results do not establish beneficial transfer to downstream paralinguistic tasks.
Determining when improvements in controlled acoustic perception translate into downstream gains will require broader multi-attribute training and more systematic evaluation on natural speech.

\section{Required Computational Resources}
\label{sec:computational-resources}
All experiments were run on a single compute node using one NVIDIA GH200 GPU (120 GB memory).
\par

Each experiment corresponding to a single model–category pair, including benchmarking, linear probing, and fine-tuning, was completed within four GPU hours.

\section{Licenses of the Existing Assets}
\label{sec:licenses}
All open-source models used in this work permit use for research purposes.
Kimi-Audio, MiMo-Audio, and Step-Audio~2~mini are released under the MIT License, while Audio~Flamingo~3 allows usage for non-commercial research purposes.

Our use of Gemini 3 Flash complies with the terms of the Gemini API~\footnote{\url{https://ai.google.dev/gemini-api/terms}}~\footnote{\url{https://policies.google.com/terms/generative-ai/use-policy}}.

The VCTK dataset is released under the CC BY 4.0 license.
Since VocalGrad is constructed from augmented versions of VCTK audio clips, we release VocalGrad under the same CC BY 4.0 license.

The additional datasets used in our auxiliary experiments are also used in accordance with their respective licenses.
The Loquacious Set preserves the licenses of its constituent source datasets (CC0, CC BY 3.0, or CC BY 4.0).
The Common Voice spontaneous-speech data are released under CC0 1.0, and the MELD audio distribution used in our experiments is released under GPL-3.0.

\section{Broader Impacts}
\label{sec:broader-impacts}

This work introduces a benchmark for evaluating acoustic perception in audio language models, which may contribute to more reliable and robust audio-based AI systems.
While the dataset is constructed from publicly available speech data (VCTK), potential risks may arise from the use of real human speech recordings.
To mitigate these risks, VocalGrad consists only of controlled acoustic transformations and does not introduce additional personal or sensitive information beyond the original dataset.

\section{AI Assistance Usage}
AI-assisted tools, including ChatGPT\footnote{\url{https://chatgpt.com/}}, Codex\footnote{\url{https://openai.com/codex/}}, and Google Gemini\footnote{\url{https://gemini.google.com/}}, were used for writing refinement and coding assistance. 
These tools were not used in the development of core methods or experimental design, and their usage complies with the LLM policy described in the NeurIPS handbook.

%% file: figures/human-annotation-ui.tex
\begin{figure}[t]
    \centering
    \includegraphics[width=0.8\linewidth]{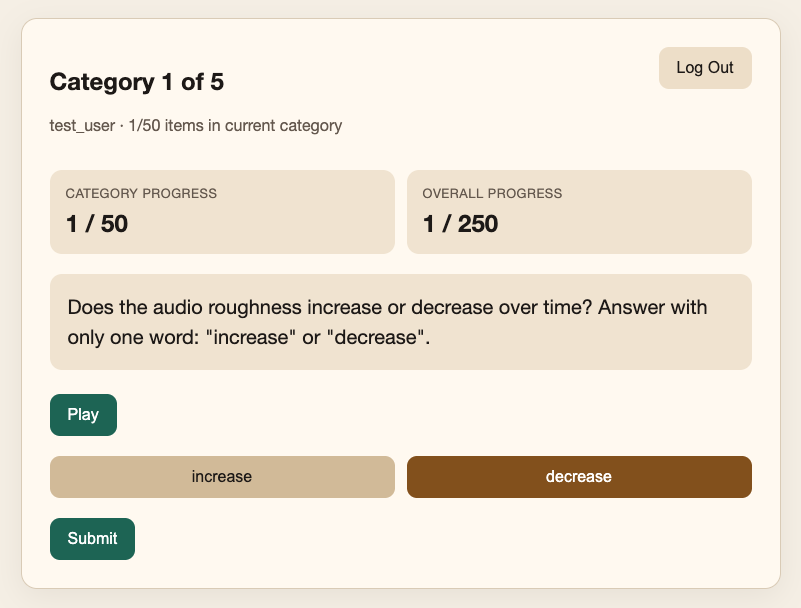}
    \caption{
        Screenshot of the web-based tool used for human annotation of VocalGrad.
    }
    \label{fig:human-annotation-ui}
\end{figure}

%% file: tables/trajectory-performance.tex
\begin{table}[t]
    \centering
    \caption{Human annotator average and model accuracies (\%) on VocalGrad by temporal trajectory.}
    \label{tab:trajectory-performance}
    \begin{tabular}{lrrrr}
        \toprule
        Model
        & \texttt{jump}
        & \texttt{linear}
        & \texttt{quad\_first\_flat}
        & \texttt{quad\_last\_flat} \\
        \midrule
        Human Avg. & 98.5 & 93.7 & 92.0 & 93.8 \\
        \midrule
        Gemini 3 Flash & 50.5 & 49.9 & 49.4 & 50.9 \\
        Kimi-Audio & 52.8 & 51.3 & 50.1 & 53.0 \\
        MiMo-Audio & 51.6 & 49.9 & 49.3 & 50.4 \\
        Step-Audio~2 mini & 50.0 & 50.0 & 49.8 & 50.0 \\
        Audio~Flamingo~3 & 49.7 & 49.6 & 49.5 & 49.7 \\
        \bottomrule
    \end{tabular}
\end{table}

%% file: figures/background_noise_voice_pitch_pair.tex
\begin{figure}[t]
\centering
\includegraphics[width=\linewidth]{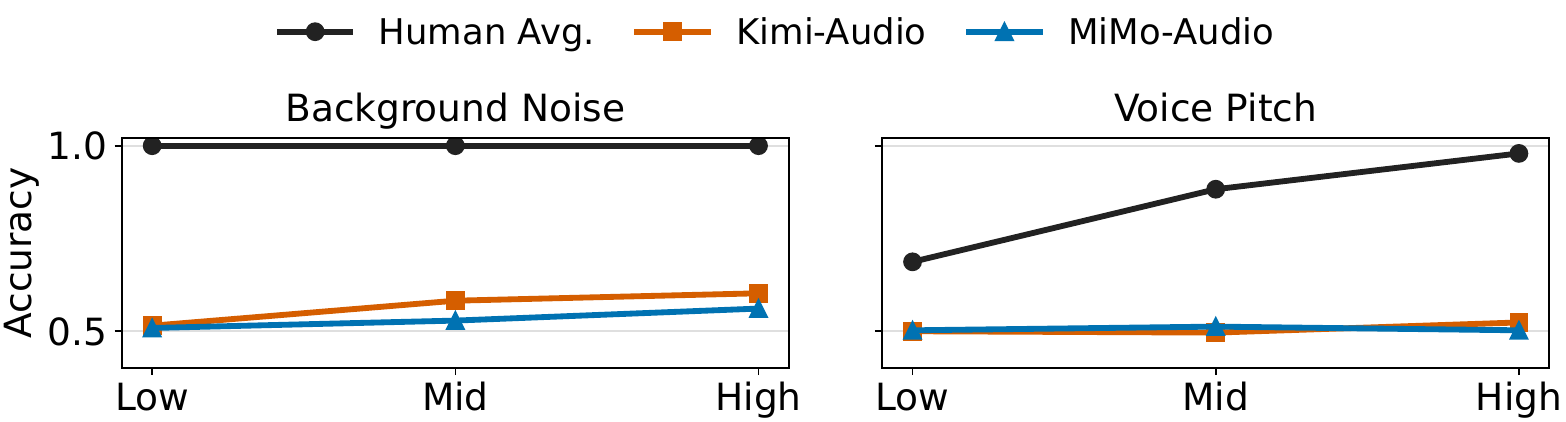}
\caption{Tier-wise accuracy for Background Noise and Voice Pitch.}
\label{fig:bgnoise-pitch-tier-accuracy}
\end{figure}

%% file: tables/parsing_output_match.tex
\begin{table}[t]
\centering
\small
\caption{Rates (\%) of raw model outputs that exactly match the target direction words or match the rule-based parser on VocalGrad. }
\label{tab:vocalgrad-parsing-output-match}
\begin{tabular}{lrr}
\toprule
Model & Exact match & Rule-based match \\
\midrule
Gemini 3 Flash & 83.9 & 99.1 \\
Kimi-Audio & 99.9 & 99.9 \\
MiMo-Audio & 99.9 & 100.0 \\
Step-Audio~2~mini & 100.0 & 100.0 \\
Audio~Flamingo~3 & 100.0 & 100.0 \\
\bottomrule
\end{tabular}
\end{table}

%% file: tables/exact-match-accuracy.tex
\begin{table}[t]
\centering
\small
\setlength{\tabcolsep}{2pt}
\caption{Model accuracies (\%) on VocalGrad when only exact raw outputs are counted as correct.}
\label{tab:vocalgrad-exact-match-accuracy}
\begin{tabular}{lrrrrrrrrrr}
\toprule
Model & Avg. & speed & pitch & volume & distortion & roughness & noise & echo & clarity & vibration \\
\midrule
Gemini 3 Flash & 42.8 & 45.8 & 50.2 & 47.6 & 27.1 & 15.3 & 52.0 & 49.0 & 52.8 & 45.4 \\
Kimi-Audio & \textbf{51.8} & \textbf{53.1} & \textbf{50.5} & 50.5 & \textbf{51.6} & 48.0 & \textbf{56.5} & \textbf{54.5} & \textbf{53.6} & 47.6 \\
MiMo-Audio & 50.3 & 50.7 & \textbf{50.5} & \textbf{52.7} & 49.8 & 43.2 & 53.2 & 50.0 & 52.2 & \textbf{50.3} \\
Step-Audio~2~mini & 50.0 & 49.9 & 50.2 & 49.9 & 50.2 & \textbf{49.9} & 49.9 & 50.0 & 49.8 & 49.8 \\
Audio~Flamingo~3 & 49.6 & 50.0 & 50.0 & 50.0 & 50.0 & 46.7 & 50.0 & 50.0 & 49.9 & 50.0 \\
\bottomrule
\end{tabular}
\end{table}

%% file: tables/category_non_exact_match.tex
\begin{table}[t]
\centering
\small
\setlength{\tabcolsep}{2pt}
\caption{Category-wise non-exact-match rates (\%) for raw model outputs on VocalGrad.}
\label{tab:vocalgrad-category-non-exact-match}
\begin{tabular}{lrrrrrrrrrr}
\toprule
Model & Avg. & speed & pitch & volume & distortion & roughness & noise & echo & clarity & vibration \\
\midrule
Gemini 3 Flash & 16.1 & 9.1 & 3.3 & 5.0 & 43.5 & 67.1 & 7.0 & 2.4 & 2.7 & 4.5 \\
Kimi-Audio & 0.1 & 0.0 & 0.0 & 0.0 & 0.0 & 0.1 & 0.0 & 0.4 & 0.0 & 0.0 \\
MiMo-Audio & 0.1 & 0.0 & 0.0 & 0.6 & 0.0 & 0.0 & 0.0 & 0.1 & 0.0 & 0.0 \\
Step-Audio~2~mini & 0.0 & 0.1 & 0.0 & 0.0 & 0.0 & 0.0 & 0.0 & 0.0 & 0.0 & 0.0 \\
Audio~Flamingo~3 & 0.0 & 0.0 & 0.0 & 0.0 & 0.0 & 0.0 & 0.0 & 0.0 & 0.0 & 0.0 \\
\bottomrule
\end{tabular}
\end{table}

%% file: figures/human_model_accuracy_ci_grid.tex
\begin{figure}[t]
    \centering
    \includegraphics[width=\linewidth]{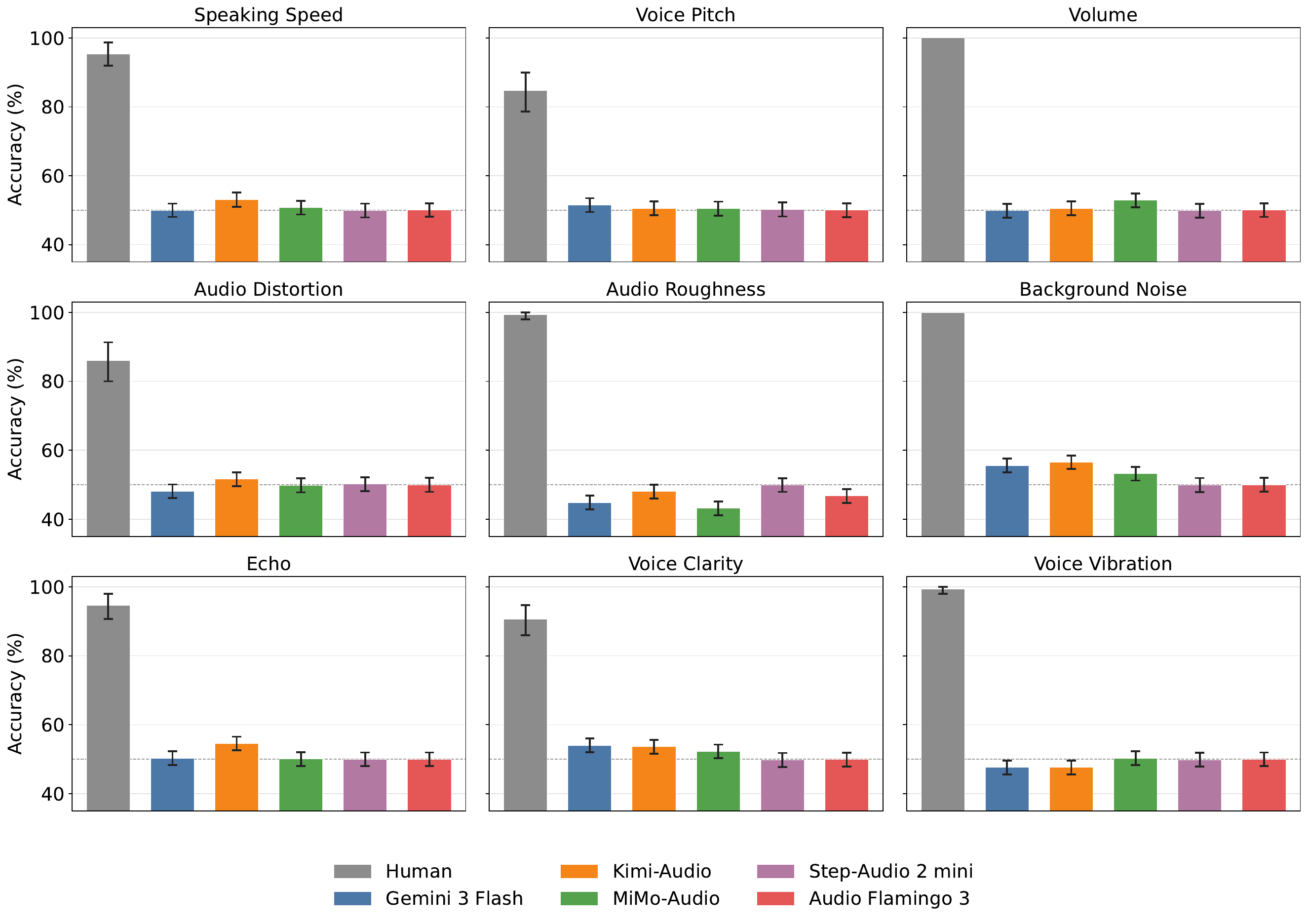}
    \caption{
    Accuracy (\%) of humans and LALMs on the VocalGrad benchmark,
    with error bars representing binomial simulation-based
    95\% confidence intervals.
    Error bars are omitted for 100\% accuracy.
    }
    \label{fig:human-model-accuracy-ci-grid}
\end{figure}

%% file: tables/original_vs_source_language_bias.tex
\begin{table}[t]
\centering
\small
\caption{Increase response bias (\%) on original VocalGrad data and source clips.}
\label{tab:original-source-increase-response}
\begin{tabular}{llrcrcr}
\toprule
Clips & Model & Avg. & Max category & Max value & Min category & Min value \\
\midrule
VocalGrad & Gemini 3 Flash & 45.8 & distortion & 85.7 & echo & 2.1 \\
 & Kimi-Audio & 24.5 & noise & 69.5 & volume & 0.8 \\
 & MiMo-Audio & 56.6 & echo & 90.9 & noise & 3.9 \\
 & Step-Audio~2~mini & 87.0 & roughness & 99.9 & echo & 0.0 \\
 & Audio~Flamingo~3 & 0.7 & roughness & 5.9 & noise & 0.0 \\
\midrule
VCTK source & Gemini 3 Flash & 47.8 & distortion & 90.7 & echo & 2.0 \\
 & Kimi-Audio & 11.1 & distortion & 41.0 & roughness & 0.0 \\
 & MiMo-Audio & 47.6 & echo & 89.9 & noise & 2.0 \\
 & Step-Audio~2~mini & 86.8 & roughness & 100.0 & echo & 0.0 \\
 & Audio~Flamingo~3 & 0.0 & volume & 0.0 & distortion & 0.0 \\
\bottomrule
\end{tabular}
\end{table}

%% file: tables/human_model_accuracy_comparison_swap_order.tex
\begin{table}[t]
\centering
\small
\setlength{\tabcolsep}{2pt}
\caption{VocalGrad benchmarking accuracy (\%) with the \texttt{swap\_order} prompt.}
\label{tab:vocalgrad-human-model-accuracy-swap-order}
\begin{tabular}{lrrrrrrrrrr}
\toprule
Model & Avg. & speed & pitch & volume & distortion & roughness & noise & echo & clarity & vibration \\
\midrule
Kimi-Audio & \textbf{51.4} & \textbf{50.7} & \textbf{50.2} & \textbf{50.1} & 49.7 & \textbf{62.0} & 47.4 & 50.0 & \textbf{52.8} & \textbf{50.1} \\
MiMo-Audio & 49.0 & 50.5 & 49.4 & 46.3 & 50.0 & 47.3 & 49.0 & 48.8 & 50.5 & 49.2 \\
Step-Audio~2~mini & 50.0 & 50.0 & 50.0 & 50.0 & 50.0 & 50.0 & 50.0 & 50.0 & 50.0 & 50.0 \\
Audio~Flamingo~3 & 50.0 & 50.0 & 50.0 & 50.0 & 50.0 & 50.0 & 50.0 & 50.0 & 50.0 & 50.0 \\
\bottomrule
\end{tabular}
\end{table}

%% file: tables/human_model_accuracy_comparison_audio_ref.tex
\begin{table}[t]
\centering
\small
\setlength{\tabcolsep}{2pt}
\caption{VocalGrad benchmarking accuracy (\%) with the \texttt{audio\_ref} prompt.}
\label{tab:vocalgrad-human-model-accuracy-audio-ref}
\begin{tabular}{lrrrrrrrrrr}
\toprule
Model & Avg. & speed & pitch & volume & distortion & roughness & noise & echo & clarity & vibration \\
\midrule
Kimi-Audio & \textbf{51.1} & \textbf{51.7} & \textbf{50.5} & 50.0 & 49.7 & \textbf{50.9} & \textbf{53.9} & \textbf{51.8} & \textbf{52.5} & 49.0 \\
MiMo-Audio & 49.8 & 50.0 & 50.2 & 49.2 & \textbf{51.4} & 44.5 & 53.0 & 50.2 & 49.9 & \textbf{50.2} \\
Step-Audio~2~mini & 50.2 & 50.2 & 50.0 & \textbf{50.1} & 50.1 & 50.2 & 50.1 & 51.0 & 50.2 & 50.1 \\
Audio~Flamingo~3 & 50.0 & 50.0 & 50.0 & 50.0 & 50.0 & 50.0 & 50.0 & 50.0 & 50.0 & 50.0 \\
\bottomrule
\end{tabular}
\end{table}

%% file: figures/vocalgrad-cross-attribute.tex
\begin{figure}[t]
  \centering
  \begin{minipage}[t]{0.49\linewidth}
    \centering
    \includegraphics[width=\linewidth]{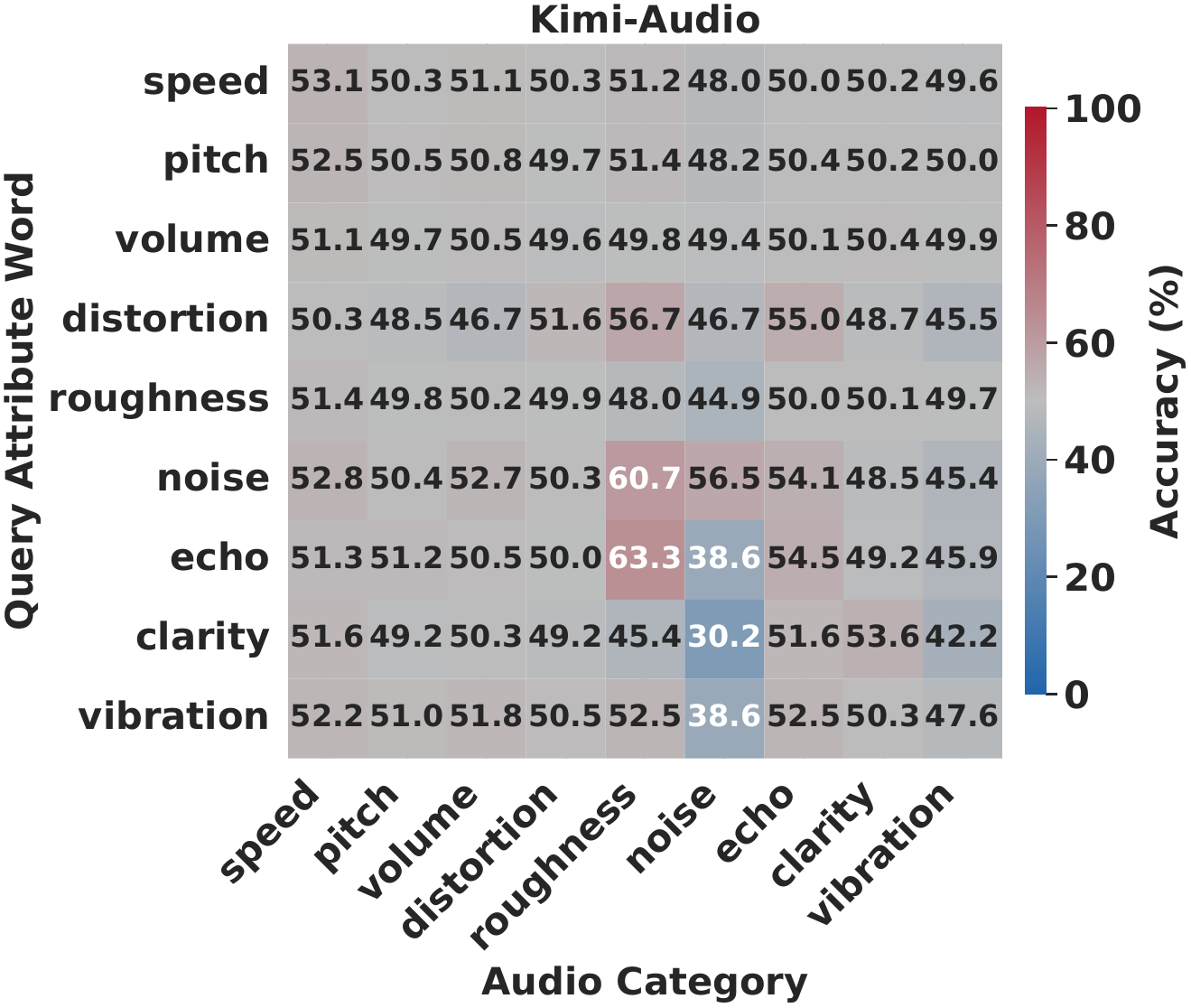}
    \vspace{-0.5em}
  \end{minipage}
  \hfill
  \begin{minipage}[t]{0.49\linewidth}
    \centering
    \includegraphics[width=\linewidth]{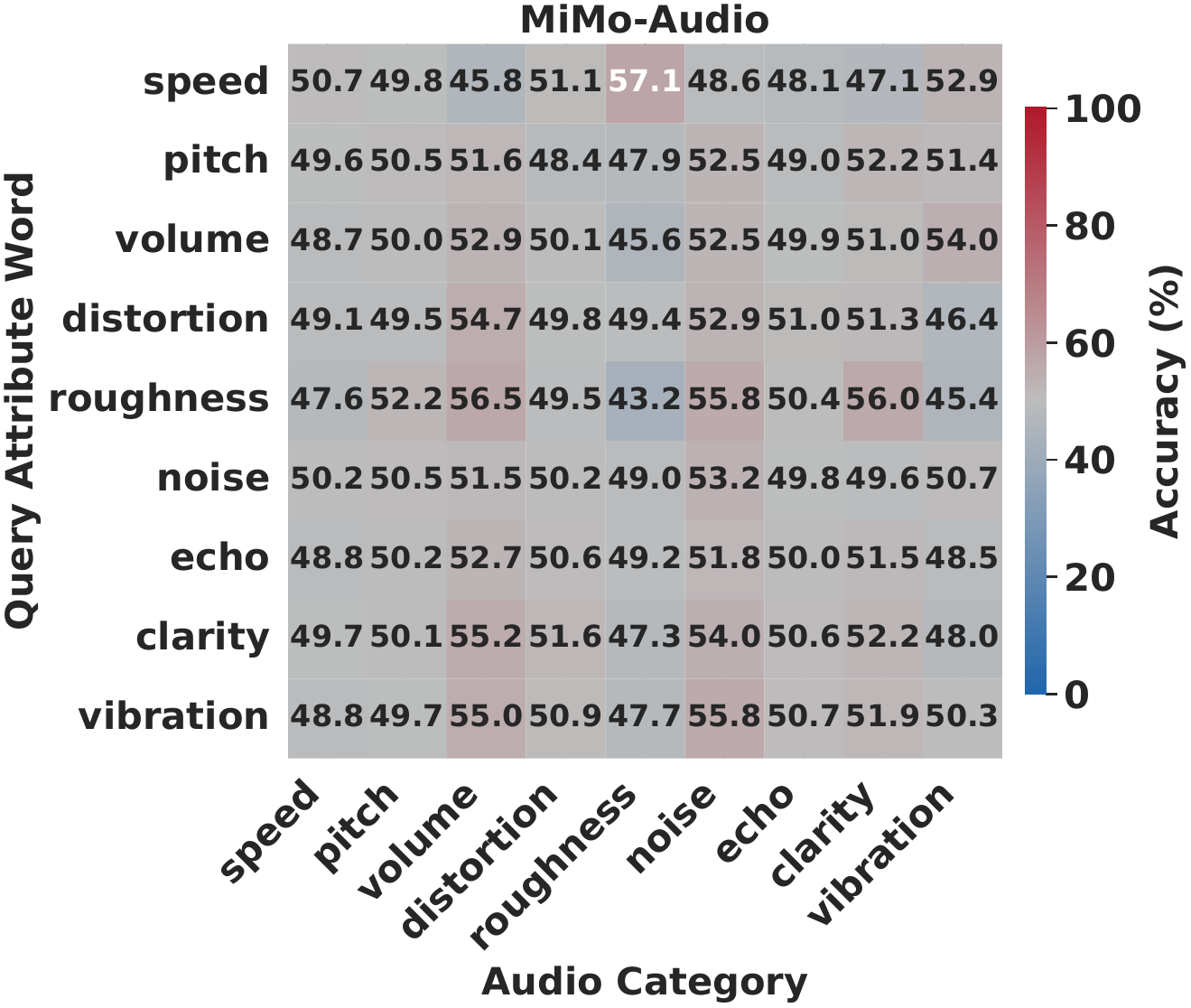}
    \vspace{-0.5em}
  \end{minipage}
  \caption{Cross-attribute VocalGrad accuracies (\%) for Kimi-Audio and MiMo-Audio. }
  \label{fig:vocalgrad-cross-attribute-kimi-mimo}
\end{figure}

%% file: tables/query-patterns.tex
\begin{table}[t]
    \centering
    \caption{Alternative category-specific query patterns used to evaluate prompt sensitivity on VocalGrad.}
    \label{tab:alternative-query-patterns}
    \small
    \begin{tabularx}{\linewidth}{llX}
        \toprule
        Category & Pattern & Query \\
        \midrule
        Volume
        & \texttt{rise\_fall}
        & Does the loudness rise or fall as the audio progresses? \\

        Volume
        & \texttt{intensity\_inc\_dec}
        & Does the sound intensity increase or decrease throughout the recording? \\

        Volume
        & \texttt{intensity\_ing}
        & Is the audio intensity increasing or decreasing over time? \\
        \midrule
        Voice Pitch
        & \texttt{tone\_rise\_fall}
        & Does the tone rise or fall over time? \\

        Voice Pitch
        & \texttt{freq\_up\_down}
        & Does the sound frequency change upward or downward over time? \\

        Voice Pitch
        & \texttt{freq\_rise\_decline}
        & Over time, does the frequency of the voice rise or decline? \\
        \midrule
        Speaking Speed
        & \texttt{pace\_fast\_slow}
        & Does the pace of speech become faster or slower over time? \\

        Speaking Speed
        & \texttt{tempo\_inc\_dec}
        & Does the speech tempo increase or decrease with time? \\

        Speaking Speed
        & \texttt{accelerate\_decelerate}
        & Does the pace of delivery accelerate or decelerate over time? \\
        \bottomrule
    \end{tabularx}
\end{table}

%% file: tables/alternative-query-performance.tex
\begin{table}[t]
    \centering
    \caption{Kimi-Audio performance with category-specific alternative query patterns on VocalGrad.
    Accuracy and the proportion of outputs mapped to the increasing direction are reported in percent.}
    \label{tab:alternative-query-performance}
    \small
    \begin{tabular}{llrr}
        \toprule
        Category & Pattern & Accuracy & Increase ratio \\
        \midrule
        Volume & \texttt{rise\_fall} & 50.1 & 0.6 \\
         & \texttt{intensity\_inc\_dec} & 50.5 & 0.8 \\
         & \texttt{intensity\_ing} & 50.0 & 0.0 \\
        \midrule
        Voice Pitch & \texttt{tone\_rise\_fall} & 50.0 & 0.1 \\
         & \texttt{freq\_up\_down} & 50.4 & 5.7 \\
         & \texttt{freq\_rise\_decline} & 51.5 & 40.1 \\
        \midrule
        Speaking Speed & \texttt{pace\_fast\_slow} & 50.7 & 0.9 \\
         & \texttt{tempo\_inc\_dec} & 51.7 & 4.2 \\
         & \texttt{accelerate\_decelerate} & 53.2 & 26.4 \\
        \bottomrule
    \end{tabular}
\end{table}

%% file: tables/vocalgrad-fewshot.tex
\begin{table}[t]
\centering
\small
\setlength{\tabcolsep}{2pt}
\caption{Few-shot prompting accuracies (\%) on VocalGrad by category.}
\label{tab:vocalgrad-fewshot}
\begin{tabular}{lrrrrrrrrrr}
\toprule
Model & Avg. & speed & pitch & volume & distortion & roughness & noise & echo & clarity & vibration \\
\midrule
Kimi-Audio & 50.2 & 51.6 & \textbf{49.3} & 51.1 & 50.0 & 49.3 & 49.2 & 50.7 & \textbf{51.0} & 49.5 \\
MiMo-Audio & 49.9 & 47.6 & 49.3 & 50.0 & \textbf{50.3} & \textbf{52.0} & 51.0 & 49.8 & 50.1 & 48.8 \\
Step-Audio~2~mini & \textbf{51.0} & \textbf{53.2} & 49.0 & \textbf{52.8} & 50.0 & 49.6 & \textbf{51.3} & \textbf{51.1} & 50.3 & \textbf{51.8} \\
\bottomrule
\end{tabular}
\end{table}

%% file: tables/source-domain-expansion.tex
\begin{table}[t]
    \centering
    \caption{Kimi-Audio accuracies (\%) on VocalGrad across different source datasets.}
    \label{tab:source-domain-expansion}
    \resizebox{\linewidth}{!}{
    \begin{tabular}{lcccccccccc}
        \toprule
        Dataset & Avg. & speed & pitch & volume & distortion
        & roughness & noise & echo & clarity & vibration \\
        \midrule
        VCTK & 51.8 & 53.1 & 50.5 & 50.5 & 51.6 & 48.0 & 56.5 & 54.5 & 53.6 & 47.6 \\
        Loquacious Set & 50.6 & 52.5 & 50.4 & 50.8 & 50.0 & 49.1 & 49.7 & 51.3 & 51.3 & 50.6 \\
        Common Voice & 52.9 & 59.0 & 54.0 & 51.2 & 50.5 & 56.1 & 50.3 & 53.7 & 53.0 & 48.4 \\
        \bottomrule
    \end{tabular}
    }
\end{table}

%% file: tables/different-problem-performance.tex
\begin{table}[t]
    \centering
    \caption{Kimi-Audio performance (\% macro-F1) on alternative VocalGrad problem formats.}
    \label{tab:alternative-problem-format-performance}
    \resizebox{\linewidth}{!}{
    \begin{tabular}{lccccccccc}
        \toprule
        Problem Format
        & speed
        & pitch
        & volume
        & distortion
        & roughness
        & noise
        & echo
        & clarity
        & vibration \\
        \midrule
        Yes/No
        & 59.4
        & 58.3
        & 47.9
        & 53.3
        & 44.9
        & 58.0
        & 45.7
        & 51.0
        & 50.4 \\
        Ternary
        & 26.5
        & 24.5
        & 24.2
        & 24.4
        & 24.1
        & 23.6
        & 21.6
        & 27.6
        & 21.8 \\
        \bottomrule
    \end{tabular}
    }
\end{table}

%% file: tables/different-problem-prediction-distribution.tex
\begin{table}[t]
    \centering
    \caption{Prediction distributions (\%) of Kimi-Audio under the binary change-detection and ternary variation tasks. Columns indicate the input clip type.}
    \label{tab:alternative-problem-format-distribution}
    \begin{tabular}{llrrr}
        \toprule
        Format & Prediction & Increasing & Decreasing & Source \\
        \midrule
        \multirow{2}{*}{Yes/No}
        & Yes & 87.6 & 85.6 & 72.8 \\
        & No  & 12.4 & 14.4 & 27.2 \\
        \midrule
        \multirow{4}{*}{Ternary}
        & Increase & 2.2  & 1.6  & 0.2 \\
        & Decrease & 92.0 & 91.1 & 89.2 \\
        & Constant & 5.6  & 7.1  & 10.3 \\
        & Unparsed & 0.2  & 0.2  & 0.2 \\
        \bottomrule
    \end{tabular}
\end{table}

%% file: tables/position-probing-query-audio.tex
\begin{table}[t]
    \centering
    \caption{
    Position-wise linear probing accuracy (\%) of Kimi-Audio under the
    Query$\rightarrow$Audio context ordering.
    $a_i$ denotes the final audio token of the $i$-th temporal segment,
    and $o$ denotes the output-prediction position.
    }
    \label{tab:position-probing-query-audio}
    \begin{tabular}{lccccc}
        \toprule
        Layer & $a_1$ & $a_2$ & $a_3$ & $a_4$ & $o$ \\
        \midrule
        1  & 85.5 & 78.3 & 78.5 & 92.4 & 93.1 \\
        10 & 82.8 & 69.6 & 69.5 & 89.8 & 88.8 \\
        25 & 75.1 & 65.7 & 64.7 & 80.5 & 85.5 \\
        \bottomrule
    \end{tabular}
\end{table}

%% file: tables/position-probing-audio-query.tex
\begin{table}[t]
    \centering
    \caption{
    Position-wise linear probing accuracy (\%) of Kimi-Audio under the
    Audio$\rightarrow$Query context ordering.
    $a_i$ denotes the final audio token of the $i$-th temporal segment,
    $q_{\mathrm{attr}}$ the final token of the attribute expression,
    $q_p$ the penultimate query token, $q_l$ the final query token,
    and $o$ the output-prediction position.
    }
    \label{tab:position-probing-audio-query}
    \begin{tabular}{lcccccccc}
        \toprule
        Layer
        & $a_1$
        & $a_2$
        & $a_3$
        & $a_4$
        & $q_{\mathrm{attr}}$
        & $q_p$
        & $q_l$
        & $o$ \\
        \midrule
        1
        & 86.5
        & 80.3
        & 80.6
        & 92.9
        & 91.6
        & 87.4
        & 89.7
        & 90.6 \\
        10
        & 81.1
        & 66.6
        & 66.7
        & 84.8
        & 80.6
        & 79.3
        & 84.6
        & 84.7 \\
        25
        & 74.1
        & 59.9
        & 59.9
        & 70.6
        & 71.1
        & 78.2
        & 78.7
        & 76.8 \\
        \bottomrule
    \end{tabular}
\end{table}

%% file: tables/finetune-all-results.tex
\begin{table}[t]
\centering
\small
\setlength{\tabcolsep}{2pt}
\caption{Performance (\%) of all fine-tuned models on VocalGrad.}
\label{tab:finetune-all-results}
\begin{tabular}{llrrrrrrrrr}
\toprule
Method & Model & speed & pitch & volume & distortion & roughness & noise & echo & clarity & vibration \\
\midrule
LM-Head & Kimi-Audio & \textbf{73.2} & \textbf{57.0} & \textbf{88.2} & 52.2 & \textbf{97.7} & \textbf{98.6} & \textbf{61.5} & 57.4 & 53.3 \\
 & MiMo-Audio & 53.8 & 50.4 & 85.2 & \textbf{68.0} & 83.8 & 72.3 & 52.9 & \textbf{78.1} & \textbf{77.5} \\
 & Step-Audio~2~mini & 56.7 & 50.2 & 52.9 & 51.7 & 66.7 & 76.4 & 50.0 & 50.5 & 50.8 \\
 & Audio~Flamingo~3 & 58.9 & 55.6 & 69.3 & 51.3 & 84.2 & 78.9 & 51.6 & 55.2 & 58.8 \\
\midrule
All-Linear & Kimi-Audio & \textbf{99.0} & 98.7 & \textbf{100.0} & 88.5 & \textbf{100.0} & \textbf{100.0} & 96.8 & 87.0 & 99.2 \\
 & MiMo-Audio & 98.4 & 97.2 & \textbf{100.0} & \textbf{99.1} & \textbf{100.0} & \textbf{100.0} & 94.6 & \textbf{100.0} & 99.7 \\
 & Step-Audio~2~mini & \textbf{99.0} & \textbf{99.6} & \textbf{100.0} & 95.1 & \textbf{100.0} & \textbf{100.0} & 99.7 & 99.6 & 99.8 \\
 & Audio~Flamingo~3 & 98.3 & 99.5 & \textbf{100.0} & 97.2 & 100.0 & \textbf{100.0} & \textbf{100.0} & 99.6 & \textbf{100.0} \\
\bottomrule
\end{tabular}
\end{table}

%% file: tables/fine-tuning-category-alignment.tex
\begin{table}[t]
\centering
\small
\caption{Mean signed cosine similarity and mean absolute cosine
similarity between category-specific increase--decrease directions
in Kimi-Audio hidden representations. Each cell reports signed / absolute cosine similarity, averaged over the 36 unordered pairs of the nine augmentation categories.}
\label{tab:category-direction-cosine}
\begin{tabular}{rccc}
\toprule
Layer index & Base & Volume-tuned & Voice-Pitch-tuned \\
\midrule
5  & 0.005 / 0.148 &  0.004 / 0.296 & 0.019 / 0.146 \\
10 & 0.020 / 0.153 & -0.059 / 0.490 & 0.028 / 0.187 \\
15 & 0.023 / 0.132 & -0.082 / 0.586 & 0.038 / 0.231 \\
20 & 0.079 / 0.190 & -0.089 / 0.599 & 0.139 / 0.709 \\
25 & 0.094 / 0.234 & -0.084 / 0.657 & 0.134 / 0.737 \\
28 & 0.054 / 0.205 & -0.095 / 0.729 & 0.116 / 0.682 \\
\bottomrule
\end{tabular}
\end{table}

%% file: figures/fine-tuned-model-similarity-matrix.tex
\begin{figure}[t]
    \centering
    \includegraphics[width=\linewidth]{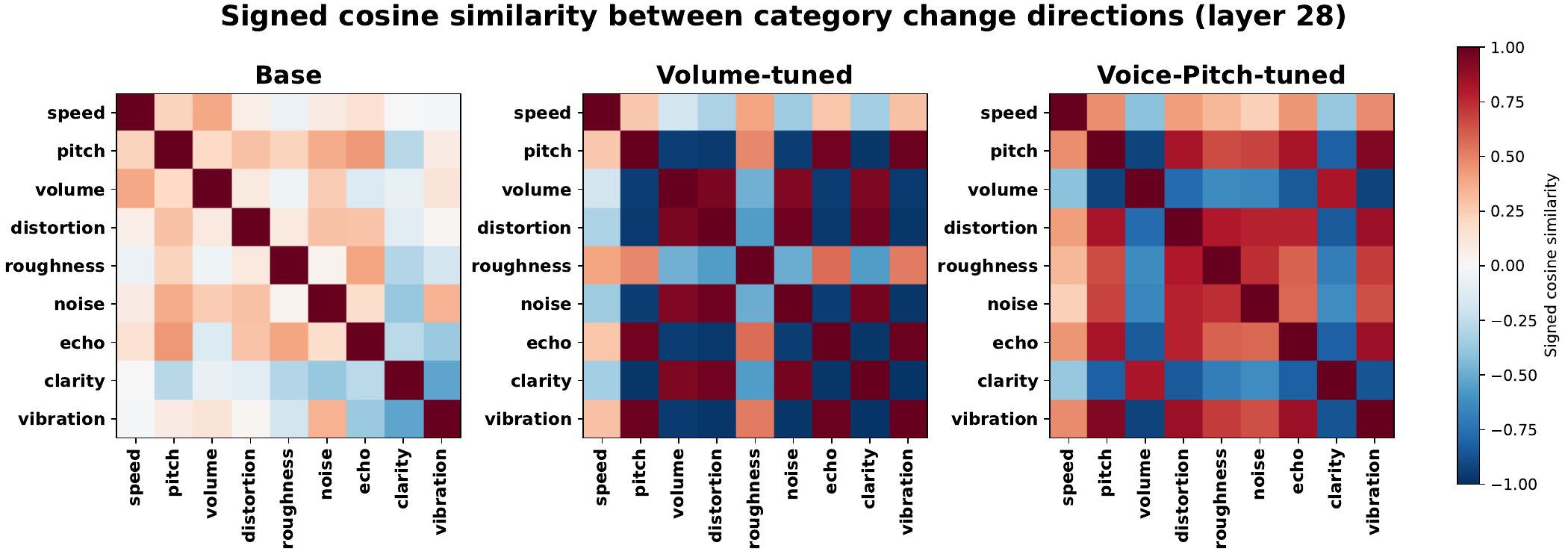}
    \caption{
        Signed cosine similarity between category-specific increase--decrease
        directions in Kimi-Audio hidden representations at layer 28.
        Results are shown for the base, Volume-tuned, and Voice-Pitch-tuned models.
    }
    \label{fig:fine-tuned-model-similarity-matrix}
\end{figure}

%% file: tables/meld-generalization.tex
\begin{table}[t]
    \centering
    \caption{
    Performance (\%) of the base and VocalGrad-fine-tuned Kimi-Audio models
    on the MELD emotion recognition dataset.
    The best result among the three models for each metric and category is shown in bold.
    }
    \label{tab:meld-transfer}
    \small

    \textbf{Precision}
    \vspace{2pt}

    \resizebox{\linewidth}{!}{
    \begin{tabular}{lrrrrrrrr}
\toprule
Model & Macro & Anger & Disgust & Fear & Joy & Neutral & Sadness & Surprise \\
\midrule
Base & \textbf{41.7} & \textbf{46.7} & \textbf{17.6} & \textbf{16.5} & \textbf{46.6} & 67.2 & 49.3 & \textbf{48.3} \\
Volume-tuned & 40.7 & 46.6 & 17.4 & 15.2 & 44.3 & 62.5 & \textbf{51.8} & 47.3 \\
Voice-Pitch-tuned & 35.3 & 41.5 & 14.9 & 11.7 & 31.9 & \textbf{67.6} & 42.0 & 37.5 \\
\midrule
\multicolumn{9}{c}{\textbf{Recall}} \\
\midrule
Model & Macro & Anger & Disgust & Fear & Joy & Neutral & Sadness & Surprise \\
\midrule
Base & \textbf{42.7} & \textbf{47.2} & \textbf{30.9} & \textbf{30.0} & 52.7 & 65.7 & \textbf{32.7} & \textbf{39.5} \\
Volume-tuned & 37.2 & 38.3 & 29.4 & 20.0 & 44.5 & \textbf{72.9} & 27.4 & 28.1 \\
Voice-Pitch-tuned & 37.0 & 34.2 & \textbf{30.9} & 14.0 & \textbf{68.9} & 44.6 & 27.9 & 38.8 \\
\midrule
\multicolumn{9}{c}{\textbf{F1}} \\
\midrule
Model & Macro & Anger & Disgust & Fear & Joy & Neutral & Sadness & Surprise \\
\midrule
Base & \textbf{41.3} & \textbf{47.0} & \textbf{22.5} & \textbf{21.3} & \textbf{49.5} & 66.4 & \textbf{39.3} & \textbf{43.4} \\
Volume-tuned & 37.7 & 42.0 & 21.9 & 17.2 & 44.4 & \textbf{67.3} & 35.8 & 35.3 \\
Voice-Pitch-tuned & 34.2 & 37.5 & 20.1 & 12.7 & 43.6 & 53.7 & 33.5 & 38.1 \\
\bottomrule
    \end{tabular}
    }
\end{table}

%% file: checklist.tex
\section*{NeurIPS Paper Checklist}

\begin{enumerate}

\item {\bf Claims}
    \item[] Question: Do the main claims made in the abstract and introduction accurately reflect the paper's contributions and scope?
    \item[] Answer: \answerYes{}
    \item[] Justification: The claims in the abstract and introduction accurately reflect the contributions presented in the paper. These include the introduction of a controlled benchmark for evaluating acoustic perception in LALMs, empirical evaluation of state-of-the-art models, identification of bottlenecks in acoustic perception, and validation of performance improvements through fine-tuning.

\item {\bf Limitations}
    \item[] Question: Does the paper discuss the limitations of the work performed by the authors?
    \item[] Answer: \answerYes{}
    \item[] Justification: The paper includes a dedicated limitations section (Appendix~\ref{sec:limitations}) that discusses three main limitations: (1) limited generalization from controlled synthetic transformations to naturally occurring, co-varying acoustic changes, (2) fixed temporal resolution in the linear probing setup, and (3) limited analysis of category-level differences in information retention across LALM representations.

\item {\bf Theory assumptions and proofs}
    \item[] Question: For each theoretical result, does the paper provide the full set of assumptions and a complete (and correct) proof?
    \item[] Answer: \answerNA{}
    \item[] Justification: The paper does not include any theoretical results, such as formal theorems or proofs.

    \item {\bf Experimental result reproducibility}
    \item[] Question: Does the paper fully disclose all the information needed to reproduce the main experimental results of the paper to the extent that it affects the main claims and/or conclusions of the paper (regardless of whether the code and data are provided or not)?
    \item[] Answer: \answerYes{}
    \item[] Justification: We provide detailed descriptions of all experimental components, including dataset construction (Section~\ref{sec:dataset_construction}, Appendix~\ref{sec:dataset-hyperparameters}), benchmarking settings (Section~\ref{sec:evaluation-settings}, Appendix~\ref{sec:model-runtime-parameter}), linear probing (Section~\ref{sec:linear-probe-method}, Appendix~\ref{sec:model-runtime-parameter}), and LoRA fine-tuning (Section~\ref{sec:fine-tuning}, Appendix~\ref{sec:model-runtime-parameter}). These details are sufficient to reproduce the main experimental results. In addition, we provide the code used in our experiments to further support reproducibility.

\item {\bf Open access to data and code}
    \item[] Question: Does the paper provide open access to the data and code, with sufficient instructions to faithfully reproduce the main experimental results, as described in supplemental material?
    \item[] Answer: \answerYes{}
    \item[] Justification: We provide open access to the VocalGrad dataset and code at \url{https://huggingface.co/datasets/ynklab/vocalgrad} and \url{https://github.com/ynklab/vocalgrad}, respectively. These resources include documentation and scripts necessary to reproduce the main experimental results described in the paper. 

\item {\bf Experimental setting/details}
    \item[] Question: Does the paper specify all the training and test details (e.g., data splits, hyperparameters, how they were chosen, type of optimizer) necessary to understand the results?
    \item[] Answer: \answerYes{}
    \item[] Justification: We provide detailed descriptions of all experimental settings, including dataset construction (Section~\ref{sec:dataset_construction}, Appendix~\ref{sec:dataset-hyperparameters}), benchmarking settings (Section~\ref{sec:evaluation-settings}, Appendix~\ref{sec:model-runtime-parameter}), linear probing (Section~\ref{sec:linear-probe-method}, Appendix~\ref{sec:model-runtime-parameter}), and LoRA fine-tuning (Section~\ref{sec:fine-tuning}, Appendix~\ref{sec:model-runtime-parameter}). These descriptions include data splits, hyperparameters, and optimization settings, which are sufficient to understand and interpret the experimental results.

\item {\bf Experiment statistical significance}
    \item[] Question: Does the paper report error bars suitably and correctly defined or other appropriate information about the statistical significance of the experiments?
    \item[] Answer: \answerYes{}
    \item[] Justification: We report binomial simulation-based 95\% confidence intervals for benchmarking results in Appendix~\ref{sec:statistical-significance}, quantifying variability due to finite sample sizes.

\item {\bf Experiments compute resources}
    \item[] Question: For each experiment, does the paper provide sufficient information on the computer resources (type of compute workers, memory, time of execution) needed to reproduce the experiments?
    \item[] Answer: \answerYes{}
    \item[] Justification: We provide detailed information on computational resources in Appendix~\ref{sec:computational-resources}, including the hardware setup (NVIDIA GH200 GPU), memory specifications, and execution time per experiment (approximately four GPU hours per model–category pair), enabling reproducibility of the reported results.
    
\item {\bf Code of ethics}
    \item[] Question: Does the research conducted in the paper conform, in every respect, with the NeurIPS Code of Ethics \url{https://neurips.cc/public/EthicsGuidelines}?
    \item[] Answer: \answerYes{}
    \item[] Justification: To the best of our knowledge, this work complies with the NeurIPS Code of Ethics. The study involves low-risk human annotation on publicly available data, with voluntary participation and no use of sensitive personal information.

\item {\bf Broader impacts}
    \item[] Question: Does the paper discuss both potential positive societal impacts and negative societal impacts of the work performed?
    \item[] Answer: \answerYes{}
    \item[] Justification: We discuss both positive and negative societal impacts in Appendix~\ref{sec:broader-impacts}, including benefits for evaluation of 
    LALMs and potential risks related to speech data and privacy, along with mitigation strategies.
    
\item {\bf Safeguards}
    \item[] Question: Does the paper describe safeguards that have been put in place for responsible release of data or models that have a high risk for misuse (e.g., pre-trained language models, image generators, or scraped datasets)?
    \item[] Answer: \answerNA{}
    \item[] Justification: The dataset is derived from the publicly available VCTK dataset~\citep{vctk} and consists of controlled acoustic augmentations of speech recordings. The data do not include personal or sensitive information beyond what is already publicly available, and no additional risk factors are introduced through the augmentation process. Therefore, the released dataset poses minimal risk of misuse.
    Our data processing steps do not add any other risk factors such as personal data leakage.

\item {\bf Licenses for existing assets}
    \item[] Question: Are the creators or original owners of assets (e.g., code, data, models), used in the paper, properly credited and are the license and terms of use explicitly mentioned and properly respected?
    \item[] Answer: \answerYes{}
    \item[] Justification: We properly cite the original authors of all models (Gemini 3 Flash, Kimi-Audio, MiMo-Audio, Step-Audio~2~mini, and Audio~Flamingo~3), datasets (VCTK, Loquacious Set, Common Voice, MELD), and libraries used in our experiments. As described in Appendix~\ref{sec:licenses}, our use of these assets complies with their respective licenses and terms of use.

\item {\bf New assets}
    \item[] Question: Are new assets introduced in the paper well documented and is the documentation provided alongside the assets?
    \item[] Answer: \answerYes{}
    \item[] Justification: The VocalGrad dataset introduced in this work is well documented and released via a public repository. The repository includes detailed instructions on dataset structure, usage, and reproduction of experiments, along with metadata and licensing information. 

\item {\bf Crowdsourcing and research with human subjects}
    \item[] Question: For crowdsourcing experiments and research with human subjects, does the paper include the full text of instructions given to participants and screenshots, if applicable, as well as details about compensation (if any)? 
    \item[] Answer: \answerYes{}
    \item[] Justification: We provide instruction details, a screenshot of the annotation interface, and information about participant compensation in Appendix~\ref{sec:human-annotation-tool}.

\item {\bf Institutional review board (IRB) approvals or equivalent for research with human subjects}
    \item[] Question: Does the paper describe potential risks incurred by study participants, whether such risks were disclosed to the subjects, and whether Institutional Review Board (IRB) approvals (or an equivalent approval/review based on the requirements of your country or institution) were obtained?
    \item[] Answer: \answerYes{} %
    \item[] Justification: The study involves a low-risk human annotation task consisting of simple binary classification on publicly available audio data. All participants were informed about the purpose of the study and participated voluntarily, with compensation offered.
    No personal or sensitive data were collected, and only aggregated results are reported. Therefore, the work poses minimal risk to participants. Further details are provided in Appendix~\ref{sec:human-annotation-tool}.

\item {\bf Declaration of LLM usage}
    \item[] Question: Does the paper describe the usage of LLMs if it is an important, original, or non-standard component of the core methods in this research? Note that if the LLM is used only for writing, editing, or formatting purposes and does \emph{not} impact the core methodology, scientific rigor, or originality of the research, declaration is not required.
    \item[] Answer: \answerNA{}
    \item[] Justification: LLM-based tools were used for coding assistance and writing support. However, they did not contribute to the core methodology, experimental design, or results of this work.

\end{enumerate}